\documentclass[traditabstract]{aa} 
\usepackage{savesym} 
\usepackage{txfonts} 
\usepackage[english]{babel} 
\usepackage{t1enc} 
\usepackage{graphicx} 
\usepackage{subfig} 
\usepackage{pdflscape}
\usepackage{fixltx2e} 
\usepackage{natbib} 
\usepackage{footmisc}
\usepackage{url} 
\usepackage{longtable}
\usepackage{multirow}
\usepackage{setspace}
\usepackage[table]{xcolor}
\usepackage{gensymb}
\usepackage{siunitx}
\usepackage{adjustbox}
\usepackage{float}
\usepackage{booktabs}
\usepackage{indentfirst}
\usepackage{float}
\usepackage{stfloats}

\usepackage[hidelinks]{hyperref}

\hypersetup{
  colorlinks   = true, %Colours links instead of ugly boxes
  urlcolor     = blue, %Colour for external hyperlinks
  linkcolor    = blue, %Colour of internal links
  citecolor   = blue %Colour of citations
}

\RequirePackage[margin=0cm,font={color=black,small},labelfont={color=black,bf},tableposition=top,singlelinecheck=false,hypcap]{caption} % caption
\newcommand{\RNum}[1]{\uppercase\expandafter{\romannumeral #1\relax}}

\begin{document} 
 
\title{Gaps and Rings: A Near-Universal Trait of Extended Protoplanetary Discs}
\author{Q. Bosschaart\inst{1}, O.M. Guerra-Alvarado\inst{1}, N. van der Marel\inst{1}, G.D. Mulders\inst{2}}

\institute{Leiden Observatory, Leiden University, PO Box 9513, 2300 RA Leiden, the Netherlands
\\  e-mail: {\tt quincybosschaart@gmail.com}
\and Instituto de Astrof\'isica, Pontificia Universidad Cat\'olica de Chile, Av. Vicu\~na Mackenna 4860, 7820436 Macul, Santiago, Chile}

\date{Received ; accepted } 
 
\abstract{\textit{Context}. Substructures such as rings, gaps, and cavities are commonly observed in protoplanetary discs and are thought to play a key role in dust evolution and planet formation. However, a fraction of the extended discs (68\% dust radii >30 AU) in nearby star-forming regions remain unresolved, leaving their substructure content uncertain and thereby limiting our understanding of dust evolution and the initial conditions for planet formation across the full disc population.\\
\textit{Aims}. We aim to investigate the presence of substructures in previously unresolved, extended discs to assess whether all extended protoplanetary discs in the Solar neighbourhood exhibit substructures. This enables a statistical evaluation of substructure occurrence among protoplanetary discs and provides statistical constraints for disc evolution models and comparisons with exoplanet populations.\\
\textit{Methods}. We present new high-resolution ($\sim$0.12") ALMA (Atacama Large Millimeter/submillimeter Array) Band 6 continuum observations at 1.33 mm of 26 previously unresolved, extended discs within 200 pc. This completes the sample of high-resolution observations of extended discs in the nearby star-forming regions Taurus, Ophiuchus, Chamaeleon, Lupus, Upper Scorpius, Upper
Centaurus-Lupus and Lower Centaurus-Crux. We analyse radial intensity profiles using \texttt{Frankenstein} and \texttt{Galario} to detect substructures.\\
\textit{Results}. Seventeen discs show clear substructures, while nine appear compact and structureless, smooth or ambiguous due to inclination or possible binarity/late-stage infall. We detect $^{12}$CO J=2-1 emission in 15 discs, where extended CO emission is observed in four discs. Combined with literature data, our complete sample of 730 protoplanetary discs reveals that nearly all extended discs exhibit substructures, $\sim$91\% detected in the full sample, and up to $\sim$98\% when correcting for high-inclination systems where substructures may be hidden.\\
\textit{Conclusions}. Substructures are a near-universal feature of extended protoplanetary discs. Substructures are more commonly detected and proposed to be more prevalent in larger, massive discs and around higher-mass stars, and structured discs retain their dust mass over time. This is consistent with the hypothesis that dust traps, possibly induced by giant planets, are key in shaping the dust disc morphologies.
}

\keywords{} 

\titlerunning{Gaps and Rings in Extended Protoplanetary Discs}
\maketitle 
 
\section{Introduction}  
\label{sec:Intro}
Substructures in the dust distribution of protoplanetary discs, such as rings, gaps, cavities, and spirals, are frequently observed in bright, large discs, particularly those resolved with high-resolution ALMA (Atacama Large Millimeter/submillimeter Array) observations (e.g. \citealt{Andrews2018}; \citealt{Huang}; \citealt{Long}). However, this trend is shaped by observational biases: most high-resolution studies have focused on the brightest and most extended discs. In comparison, compact discs (here defined as dust discs with 68\% dust radii <30 AU) appear to show fewer substructures, as confirmed by the complete high-resolution survey of the Lupus star-forming region (\citealt{guerra2025}). These dust features are key to understanding the evolution of protoplanetary discs and the early stages of planet formation. However, \citet{Bae2023} demonstrated that substructures are more likely to be detected for discs resolved well beyond the angular resolution, which makes an analysis of dust processes across disc sizes challenging.

\indent Dust grains in protoplanetary discs undergo complex evolution; they grow through coagulation, experience fragmentation, and drift radially due to aerodynamic coupling with the gas. Radial drift, in particular, leads to the rapid inward migration of millimetre-sizes grains, posing a challenge to planet formation models that require retention of solid material at large radii (\citealt{Weidenschilling}; \citealt{Birnstiel2024}). Pressure bumps in the gas can act as dust traps, halting radial drift and enabling the accumulation of solids. These traps often result in ring-like substructures in the dust continuum, and are considered critical sites for planetesimal formation (\citealt{Pinilla2012b}; \citealt{Dullemond2018}).\\
\indent Several demographic studies of protoplanetary discs have explored the connection between disc properties and stellar mass, age, and the star-forming region (e.g. \citealt{Ansdell2016}; \citealt{Barenfeld}; \citealt{Cieza}). These surveys revealed trends such as larger and more massive discs around higher-mass stars, but lacked the angular resolution to probe the substructures. It has also been suggested that the occurrence of discs with large-scale substructures may correlate with stellar mass (\citealt{NienkeGijs}) with a similar trend for transition discs \citep{vandermarel2023}, supported by similar trends in exoplanet demographics where giant planets are more frequently found around higher-mass stars (\citealt{Johnson2010}). This similarity hints at a causal connection, where giant planets may be responsible for shaping the disc by carving rings and gaps during their formation. A growing body of evidence points to planet-disc interactions as a dominant cause, particularly for large-scale ringed structures (see e.g. \citealt{Baruteau}; \citealt{Zhang2018}), although other processes such as magnetically induced zonal flows (\citealt{Hawley}), dead zones (\citealt{Flock}), and condensation fronts (\citealt{Zhang2015}) have also been proposed.\\
\indent However, several protoplanetary discs in nearby star-forming regions remain unresolved or only marginally resolved, since generally only the brightest discs have been followed up with high-resolution observations. The study by \citet{NienkeGijs} phrased the hypothesis that discs could be divided up in structured, extended discs and smooth, compact discs, based on their findings that the structured discs retain similar millimetre flux densities when comparing discs in younger and older regions, whereas the millimetre flux densities of smooth, compact discs, the majority of the disc population, decrease over time. This would be consistent with a scenario where structured discs contain strong dust traps, whereas smooth discs do not, and radial drift reduces their observable millimetre emission over time, following dust evolution models \citep{Pinilla2020}. Gas-to-dust size ratios have been predicted to be measurements of discs with and without dust traps \citep{Toci2021}, but interpreting $^{\text{12}}$CO emission sizes as a direct tracer of the physical extent of the gas disc has been proven to be very challenging \citep{Trapman}. Recently, \citet{Pinilla2025} classified the discs in Lupus and Upper Scorpius as either structured or smooth based on visibility modelling. They concluded that since the smooth discs only change in millimetre flux but not in size as a function of evolution, the smooth discs are consistent with weak or leaky dust traps, in contrast to the strong dust traps seen in structured discs. The division in two pathways of dust evolution is also seen in e.g. Herbig discs \citep{Stapper} and reflected in the observations of sublimated H$_2$O in infrared spectra, for which radial drift has been proposed as a possible interpretation \citep[e.g.][]{Banzatti2020,Banzatti2023}. Recent JWST obserrvations reveal a substantial intrinsic scatter in H$_2$O emission properties \citep{Gasman2025}. While a comprehensive explanation is still under investigation, part of this diversity may reflect evolutionary effects as explored in recent modelling efforts \citep{Krijt2025}. Whereas caution is warranted in the interpretation of low-resolution ALMA observations \citep{Bae2023}, we build further on the hypothesis of separate evolutionary pathways in this new study with follow-up observations of the extended discs.\\
\indent In the work by \citet{NienkeGijs} a population of large discs with radii >40 AU was classified as "extended discs" that appeared smooth due to limited angular resolution. %Based on dust evolution models, s
Such discs are unlikely to remain extended without the presence of substructures that prevent radial drift, according to dust evolution models. Yet the hypothesis that these unresolved discs contain undetected rings and gaps remained untested due to the lack of high-resolution data for these extended discs.\\
\indent This study aims to address this question by analysing a new sample of 26 extended protoplanetary discs in the nearby star-forming regions Taurus, Ophiuchus, Chamaeleon, Upper Centaurus-Lupus, Upper Scorpius, and Lupus, with 68\% dust disc radii larger than 30 AU, observed with high-resolution ALMA imaging for the first time. The choice of a 30 AU threshold for extended discs is motivated by the unbiased ALMA survey of the Taurus star-forming region at 0.12" resolution by \citet{Long, Long2019}, which showed that discs with dust radii $\gtrsim$30 AU are robustly resolved and representative of the spatially extended disc population in nearby regions. We model the radial emission structure of each disc using the tools \texttt{Frankenstein} and \texttt{Galario}, identifying and characterising rings and gaps where present. By completing the high-resolution sample of extended discs within 200 pc, this work enables a statistical assessment of substructure occurrence, providing key input for models of planet formation and comparison with exoplanet population trends.\\
\indent We discuss the sample selection and observations in Section \ref{sec:Sample and Observations} and the analysis of the dust continuum observations in Section \ref{sec:Dust continuum results}. In Section \ref{sec:modelling}, we present the modelling approach for the disc substructures and the corresponding results. In Section \ref{sec:Discussion}, we discuss the dust substructures of our sample and analyse the complete sample of nearby extended protoplanetary discs, thereby answering the question whether all extended discs contain substructures. Finally, the conclusions of this work are summarised in Section \ref{sec:Conclusion}.

\section{Sample and observations}
\label{sec:Sample and Observations}
\subsection{Sample selection}
\label{subsec:sample}
The sample of analysed protoplanetary discs in this work was selected from the \citet{NienkeGijs} catalog. This catalog contains nearly 700 discs from nearby star-forming regions within a 350 pc radius, classified by observed dust morphology: 42 transition discs with inner cavities >25 AU, 37 ring discs, 474 smooth discs, and 147 non-detections. The smooth discs were further classified into extended (>40 AU disc radius (68\%)) and compact (<40 AU disc radius (68\%)) discs. Since the publication of \citet{NienkeGijs}, some extended discs have been observed at high angular resolution, revealing rings, gaps, and cavities (see e.g. \citealt{Villenave2020, Villenave2022}; \citealt{Cieza2021}). We select all discs with 68\% dust disc sizes larger than 30 AU that have not yet been observed at high angular resolutions, excluding distant discs, resulting in a sample of 23 star systems observed in the ALMA program presented in this study. This includes three binary discs, yielding a total of 26 protoplanetary discs, located in the Taurus, Ophiuchus, Chamaeleon, Upper Centaurus-Lupus, Upper Scorpius, and Lupus star-forming regions.\\
\indent The 30 AU threshold for extended discs was motivated by the unbiased ALMA disc survey of the Taurus star-forming region at 0.12" by \citet{Long, Long2019}. While this study focuses on high-resolution observations of these 26 discs, the discussion section expands to the broader population of extended discs identified in \citet{NienkeGijs} and explores their statistical properties in relation to stellar mass and other parameters. The full sample is detailed in Table \ref{tab:cleaned}. Table \ref{tab:properties} lists the host star properties, estimated dust disc sizes (column 6), and dust masses (column 7), taken from literature. For some discs, the dust sizes represent upper limits (indicated by "<").\\
\indent A discrepancy in the sample selection should be noted. SSTc2d J162738.3-235732, with a dust disc radius already known to be smaller than 14 AU, was mistakenly included as an extended disc in the ALMA proposal due to a catalog error in \citet{NienkeGijs}. SSTc2d J162738.3-243648 should have been included in the ALMA program instead. However, since SSTc2d J162738.3-235732 has now been observed at high angular resolution it will still be analysed in addition to the extended disc sample in order to accurately determine its true dust radius and assess potential substructures within this compact disc. Additionally, Haro 6-37 A and MHO 1, with radii below 30 AU, are part of binary systems where their companions (Haro 6-37 B and MHO 2) exceed the 30 AU threshold. For the third binary system, both IT Tau A and IT Tau B have dust radii near the inclusion limit (32 AU).

\subsection{Observations}
\label{subsec:Observations}
The 23 systems were observed during the ALMA Cycle 9 program (PI: G. Mulders; program ID: 2022.1.01302.S) between January 2, 2023 and June 2, 2023 in Band 6 (1.33 mm) with high spatial resolutions ($\sim$0.12"). Additional observations at a lower resolution ($\sim$0.6") were requested for short baseline coverage, but only nine discs were observed in the short baseline configuration. For each observation, the ALMA correlators were configured into four separate basebands. Two windows were allocated for continuum observations, with central frequencies at 218 and 233 GHz, each having a bandwidth of 1.875 GHz. The other two windows were designated for molecular line emission observations, centred at 220 and 230 GHz, targeting the $^{\text{12}}$CO, $^{\text{13}}$CO, and C$^{\text{18}}$O J=2-1 lines, with bandwidths of 0.234 GHz each, and a channel resolution of 0.184 km/s. For the lower angular resolution observations, On-source integration times ranged between $\sim$3 and $\sim$5 minutes per target and for the higher angular resolution observations, the integration times ranged between $\sim$10 and $\sim$30 minutes per target. The high-resolution observations of 2MASS J04154278+2909597 and 2MASS J16075796-2040087, along with the low-resolution observation of SSTc2d J163952.9-241931, were conducted over several days. Table \ref{tab:observations} summarises the observational details.\\
\indent A standard ALMA pipeline calibration was performed by the ALMA staff for both the high- and low-angular-resolution datasets. The low-resolution observations were calibrated using CASA version 6.4.1.12 with pipeline version 2022.2.0.64, while the high-resolution observations were calibrated using pipeline version 2022.2.0.68. The standard calibration included offline water vapour radiometer calibration, system temperature correction, bandpass, phase, amplitude, and flux calibrations. The images included in the data delivery were corrected for the primary beam response. Self-calibration was performed using the CASA \textit{tclean} and \textit{gaincal} tasks, starting with phase-only solutions on progressively shorter solution intervals, followed by a final round including amplitude self-calibration where the signal-to-noise ratio permitted. Typical improvements in peak signal-to-noise ranged from 20-40$\%$.

\subsubsection{Post-processing of the dust continuum data}
The final 1.3 mm dust continuum datasets were produced by averaging the calibrated visibilities over the continuum channels. Images were created using CASA's \textit{tclean} with Briggs weighting and a robust parameter of +0.5, adjusted to 0.0 or -0.5 for sources with immediately resolved substructures. We have checked the effects of frequency smearing for our binary sources and find less than 6\% differences in the flux densities when averaging over all channels compared to using a single spectral window. We consider this negligible for the purposes of this work.\\
\indent Several discs were initially off-centred due to outdated RA and Dec coordinates used for telescope pointing. To correct this, the CASA \textit{phaseshift} task was applied to align the discs. For the nine discs with both high- and low- resolution data, datasets were combined using CASA's \textit{concat} task and re-imaged with \textit{tclean}. The average beam size of the final images is $\sim0.15\times0.11$".

\subsubsection{Post-processing of the CO line emission data}
The $^{\text{12}}$CO line channel maps were extracted from the calibrated visibilities by subtracting the continuum emission from the phase-shifted UV data using CASA's \textit{uvcontsub} task with a polynomial of order 0. The line emission was cleaned using Briggs weighting with a robust parameter of +0.5. The $^{\text{12}}$CO J=2-1 transition line, with a rest frequency of 230.538 GHz, was detected in 15 discs (see column 6 in Table \ref{tab:cleaned}). For non-detections, cleaning with natural weighting was attempted as well, but this did not result in additional detections. For discs with both high- and low-resolution data available, low-resolution data was preferred for its sensitivity to large-scale structures, providing higher SNRs due to better UV coverage at shorter baselines (see column 5 in Table \ref{tab:observations}).\\
\indent Moment 0 maps were created by integrating over all velocity channels with detected emission, and moment 1 maps by calculating the intensity-weighted mean velocity over the same channels. Velocity ranges for each disc were determined from visual inspection of the channel maps. The moment maps were generated using the \texttt{bettermoments} package (\citealt{bettermoments}). The $^{\text{12}}$CO gas observations are shown and analysed in Appendix \ref{appendix:12 CO moment maps}.

\section{Dust continuum results}
\label{sec:Dust continuum results}
\subsection{Flux densities}
\begin{table*}[ht]
    \caption{Column 1 lists the source names. Column 2 provides the calculated flux densities integrated over 3RMS, while column 3 lists the total flux densities obtained from the curve-of-growth (CoG) method (mJy). Column 4 presents the flux densities from previous studies, with references in brackets (full citations below). Column 5 indicates the RMS of the background noise (mJy/beam), and column 6 gives the estimated source centre coordinates. Column 7 notes wheter $^{\text{12}}$CO gas was detected (yes/no), and column 8 lists the final disc classifications: TD (transition disc), RD (ring disc), SD (shoulder disc), HID (highly-inclined disc), CD (compact disc), and ED (extended disc). Flux densities and RMS values were derived from high-resolution images cleaned with Briggs weighting (robustness +0.5).}
    %\centering
    \hspace{-5mm}
    \begin{tabular}{l c c c c c c c c c}
        \hline\hline
        \textbf{Disc Name} & \multicolumn{3}{c}{\textbf{Flux Density}} & \textbf{RMS} & \textbf{Source Centre} & \textbf{$^{\text{12}}$CO?} & \textbf{Class}\\
        &  & \hspace{4mm}\textbf{[mJy]}& &\textbf{[mJy/beam]} & & &\\
        & \textit{F$_{3\sigma}$} & \textit{F$_{CoG}$} & Previous & & & &\\
        
        \hline
        2MASS J11095340-7634255 & 18.5 & - &- & 0.04 & 11$^{\text{h}}$09$^{\text{m}}$53$^{\text{s}}$.24 -76$^{\text{d}}$34$^{\text{m}}$25$^{\text{s}}$.61 & Yes & ED\\
        2MASS J11004022-7619280 & 27.7 & 29.9 & - & 0.04 & 11$^{\text{h}}$00$^{\text{m}}$40$^{\text{s}}$.09 -76$^{\text{d}}$19$^{\text{m}}$27$^{\text{s}}$.98 & No & SD\\
        2MASS J11111083-7641574 & 18.6 & 19.0  &- & 0.04 & 11$^{\text{h}}$11$^{\text{m}}$10$^{\text{s}}$.68 -76$^{\text{d}}$41$^{\text{m}}$57$^{\text{s}}$.25 & Yes & SD\\
        2MASS J11160287-7624533 & 4.2 & 4.5  & - & 0.03 & 11$^{\text{h}}$16$^{\text{m}}$02$^{\text{s}}$.69 -76$^{\text{d}}$24$^{\text{m}}$53$^{\text{s}}$.32 & No & RD\\
        2MASS J11104959-7717517 & 20.0 & 20.9 &- & 0.04 & 11$^{\text{h}}$10$^{\text{m}}$49$^{\text{s}}$.41 -77$^{\text{d}}$17$^{\text{m}}$51$^{\text{s}}$.72 & Yes & RD\\
        2MASS J11094742-7726290 & 49.4 & 50.4  & - & 0.04 & 11$^{\text{h}}$09$^{\text{m}}$47$^{\text{s}}$.22 -77$^{\text{d}}$26$^{\text{m}}$29$^{\text{s}}$.18 & Yes & RD\\
        SZ 133 & 25.6 & 25.2 & 27.0 (a) & 0.04 &16$^{\text{h}}$03$^{\text{m}}$29$^{\text{s}}$.37 -41$^{\text{d}}$40$^{\text{m}}$02$^{\text{s}}$.35 & Yes & HID\\
        RX J1556.1-3655 & 19.6 & 19.8 & 23.5 (a) & 0.04 & 15$^{\text{h}}$56$^{\text{m}}$02$^{\text{s}}$.07 -36$^{\text{d}}$55$^{\text{m}}$28$^{\text{s}}$.81 & No & RD\\
        SSTc2d J162652.0-243039 & 6.7 & 6.8 &10.3 (b)& 0.04 & 16$^{\text{h}}$26$^{\text{m}}$51$^{\text{s}}$.96 -24$^{\text{d}}$30$^{\text{m}}$40$^{\text{s}}$.21 & Yes & CD\\
        SSTc2d J162546.6-242336 & 21.4 & 22.6 &22.1 (b)& 0.04 & 16$^{\text{h}}$25$^{\text{m}}$46$^{\text{s}}$.63 -24$^{\text{d}}$23$^{\text{m}}$36$^{\text{s}}$.71 & Yes & HID\\
        SSTc2d J162718.4-243915 & 26.2 & 26.0 &29.0 (b)& 0.04 & 16$^{\text{h}}$27$^{\text{m}}$18$^{\text{s}}$.37 -24$^{\text{d}}$39$^{\text{m}}$15$^{\text{s}}$.39 & Yes & RD\\
        SSTc2d J162738.3-235732 & 4.1 & 4.2 & 5.3 (b) & 0.04 & 16$^{\text{h}}$27$^{\text{m}}$38$^{\text{s}}$.32 -23$^{\text{d}}$57$^{\text{m}}$33$^{\text{s}}$.13 & No & CD\\
        SSTc2d J162145.1-234232 & 37.7 & 38.4 & 40.1 (b)& 0.05 & 16$^{\text{h}}$21$^{\text{m}}$45$^{\text{s}}$.12 -23$^{\text{d}}$42$^{\text{m}}$32$^{\text{s}}$.33 & Yes & RD\\
        SSTc2d J162823.3-242241 & 19.8 & 17.7 & 17.0 (b)& 0.06 & 16$^{\text{h}}$28$^{\text{m}}$23$^{\text{s}}$.33 -24$^{\text{d}}$22$^{\text{m}}$41$^{\text{s}}$.24 & Yes & TD\\
        SSTc2d J163952.9-241931 & 6.1 & 5.9 & 7.9 (b)& 0.05 & 16$^{\text{h}}$39$^{\text{m}}$52$^{\text{s}}$.91 -24$^{\text{d}}$19$^{\text{m}}$31$^{\text{s}}$.74 & Yes & TD\\
        Haro 6-37 A & 1.4 & 1.3 & 2.1 (c)& 0.04 & 04$^{\text{h}}$46$^{\text{m}}$58$^{\text{s}}$.98 +17$^{\text{d}}$02$^{\text{m}}$37$^{\text{s}}$.59 & No & CD\\
        Haro 6-37 B & 83.2 & 84.8 & 38.5 (c)& 0.04 & 04$^{\text{h}}$46$^{\text{m}}$59$^{\text{s}}$.10 +17$^{\text{d}}$02$^{\text{m}}$39$^{\text{s}}$.63 & No & SD\\
        MHO 1 & 188.7 & 203.0 & 216.0 (c)& 0.24 & 04$^{\text{h}}$14$^{\text{m}}$26$^{\text{s}}$.28 +28$^{\text{d}}$06$^{\text{m}}$02$^{\text{s}}$.65 & Yes & SD\\
        MHO 2 & 115.0 & 127.5 & 133.3 (c)& 0.24 & 04$^{\text{h}}$14$^{\text{m}}$26$^{\text{s}}$.42 +28$^{\text{d}}$05$^{\text{m}}$59$^{\text{s}}$.02 & Yes & TD\\
        IT Tau A & 6.6 & 4.8 & 7.0 (c)& 0.04 & 04$^{\text{h}}$33$^{\text{m}}$54$^{\text{s}}$.73 +26$^{\text{d}}$13$^{\text{m}}$27$^{\text{s}}$.11 & No & CD\\
        IT Tau B & 3.4 & 3.4 & 4.2 (c)& 0.04 & 04$^{\text{h}}$33$^{\text{m}}$54$^{\text{s}}$.60 +26$^{\text{d}}$13$^{\text{m}}$25$^{\text{s}}$.38 & No & CD\\
        SU Aur & 17.4 & 17.6 & 27.4 (c)& 0.06 & 04$^{\text{h}}$55$^{\text{m}}$59$^{\text{s}}$.39 +30$^{\text{d}}$34$^{\text{m}}$00$^{\text{s}}$.93 & Yes & TD\\
        CY Tau & 88.0 & 90.5 & 79.4 (c)& 0.07 & 04$^{\text{h}}$17$^{\text{m}}$33$^{\text{s}}$.74 +28$^{\text{d}}$20$^{\text{m}}$46$^{\text{s}}$.21 & No & SD\\
        2MASS J04154278+2909597 & 14.6 & 14.3 & 19.9 (c)& 0.03 & 04$^{\text{h}}$15$^{\text{m}}$42$^{\text{s}}$.81 +29$^{\text{d}}$09$^{\text{m}}$59$^{\text{s}}$.40 & No & TD\\
        HD 139614 & 167.3 & 164.2 & 197.2 (d)& 0.06 & 15$^{\text{h}}$40$^{\text{m}}$46$^{\text{s}}$.35 -42$^{\text{d}}$29$^{\text{m}}$54$^{\text{s}}$.12 & Yes & TD\\
        2MASS J16075796-2040087 & 9.3 & 9.4 & - & 0.04 & 16$^{\text{h}}$07$^{\text{m}}$57$^{\text{s}}$.95 -20$^{\text{d}}$40$^{\text{m}}$09$^{\text{s}}$.29 & No & CD\\

        \hline
    \end{tabular}
    \caption*{\textbf{References.} (a) \citet{Ansdell}, (b) \citet{Cieza}, (c) \citet{Akeson}, (d) \citet{Stapper}}
\label{tab:cleaned}
\end{table*}
The flux densities, root-mean-square (RMS) of the background noise, and refined source centre coordinates of the 26 protoplanetary discs are summarised in Table \ref{tab:cleaned}. For comparison, previously published Band 6 flux densities are also included. We calculated two types of flux densities for each disc. First the flux density integrated above a 3$\sigma$ threshold (\textit{F$_{3\sigma}$}), with $\sigma$ the RMS of the background noise. Second, the total flux density derived from the curve-of-growth (CoG) method (\textit{F$_{CoG}$}), corresponding to the asymptotic value where the CoG curve flattens (see Section \ref{subsec:modelling_results}). Both approaches give similar values in the total flux density.
For consistency, these values were calculated from the high angular resolution images, cleaned using a Briggs weighting scheme with a robustness parameter of +0.5. The derived flux densities range between 1.3 and 203.0 mJy. The RMS noise for most discs falls within the range between 0.03 and 0.07 mJy/beam, with the exception of the two discs in the MHO 1+2 system, which exhibit an elevated RMS of 0.24 mJy/beam.\\
\indent Our calculated flux densities generally align with previous measurements, within the expected 10$\%$ ALMA Band 6 calibration uncertainty, except for Haro 6-37 B. This discrepancy may stem from the lower RMS and image fidelity in \citet{Akeson}, as their ALMA Cycle 2 observations were taken during ALMA Early Science with an incomplete array. Differences in central frequency also likely contribute to the flux variation. \\
\indent Small discrepancies between measurements can stem from differences in calibration, data reduction, and cleaning techniques. We used a uniform approach in calculating the flux densities across our dataset. Given the general agreement with prior studies within 10$\%$ flux calibration uncertainty and our consistent methodology, we can conclude that the presented flux densities are robust.

\subsection{Disc features}
\begin{figure*}[hbt]
    %\hspace{-12mm}
    %\centering
    \includegraphics[scale=0.21]{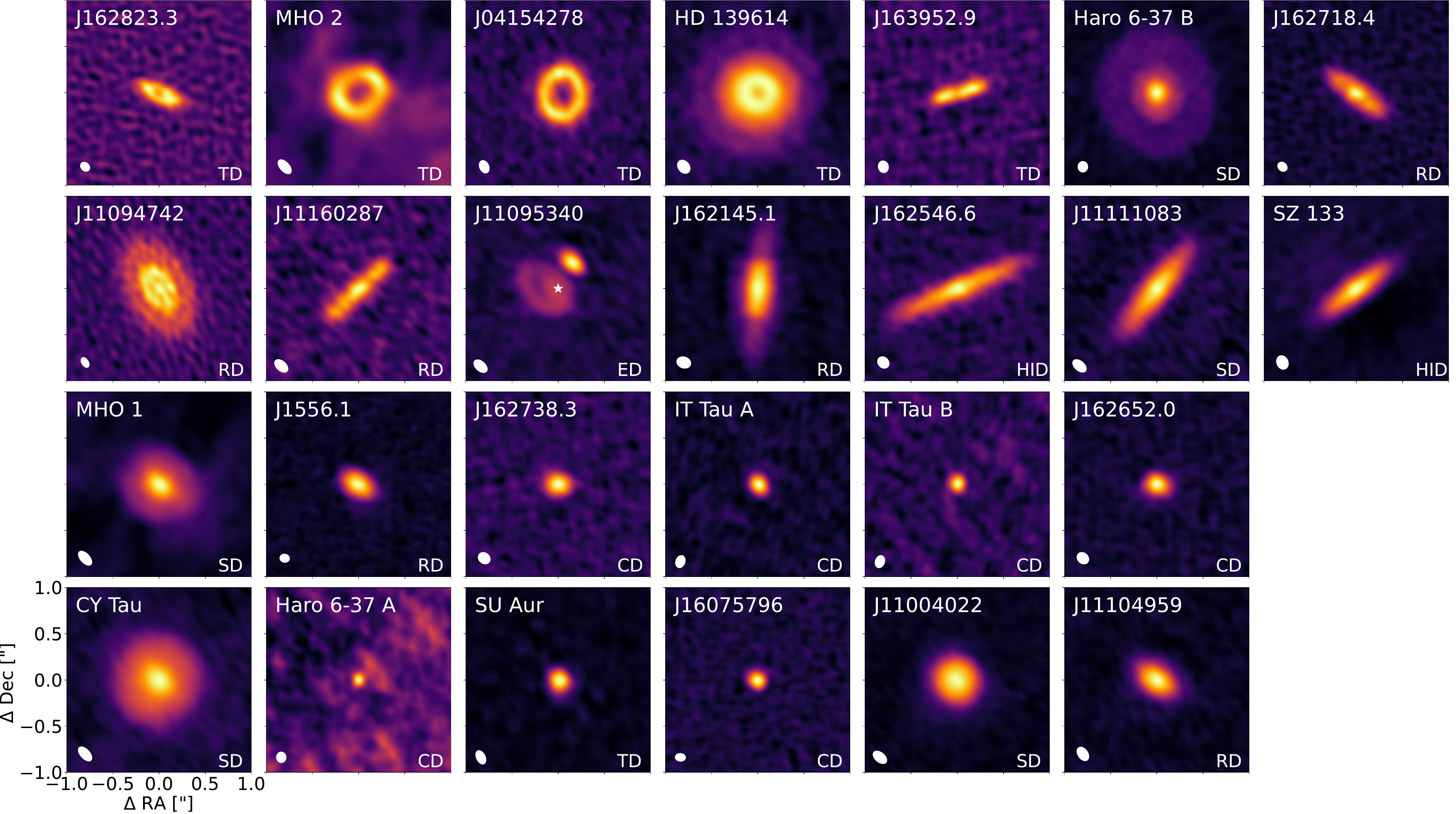}
    \caption{Gallery of the 1.3 mm cleaned continuum emission images for the full sample. All images were cleaned using a Briggs weighting scheme with a robust value of +0.5, except for SSTc2d J163952.9-241931 and SSTc2d J162718.4-243915 (robust = 0.0), and SSTc2d J162823.3-242241 and 2MASS J11094742-7726290 (robust = -0.5). RA and Dec offsets from the disc centre are shown in arcseconds on the x- and y-axes, respectively. To emphasise the weak outer emission of a few discs, an asinh scaling function was applied. Each panel spans $2.0\text{"}\times2.0\text{"}$, with white ellipses in the lower left corner indicating the beam sizes. The colour scale represents the intensity in Jy/beam. The final disc classifications are marked in the lower right corner: TD (transition disc), RD (ring disc), SD (shoulder disc), HID (highly-inclined disc), CD (compact disc), ED (extended disc). For discs with names starting with SSTc2d, 2MASS, or RX, the names in the upper left corners are abbreviated to the first part after "J" (e.g., SSTc2d J162823.3-242241 is labelled as J162823.3). The white star in 2MASS J11095340-7634255 marks the current position of the central star.}
    \label{fig:clean}
\end{figure*}
Figure \ref{fig:clean} presents the 1.3 mm cleaned continuum images of the 26 sources. For the nine discs with both long and short baseline configuration data (see Table \ref{tab:observations}) %(2MASS J11095340-7634255, 2MASS J11004022-7619280, 2MASS J11111083-7641574, 2MASS J11160287-7624533, 2MASS J11104959-7717517, 2MASS J11094742-7726290, SZ 133, SSTc2d J163952.9-241931, and HD 139614), 
concatenated images are shown. For the remaining discs, only the high-resolution images are displayed. Visual inspection of the cleaned images immediately reveals that some discs exhibit flux deficits, indicating the presence of gaps and cavities within these discs. Based on these images, we can visually classify the discs into four distinct categories: transition discs, ring discs, highly-inclined discs, and discs lacking visual substructures, as described below.\\
\indent Transition discs are characterised by a depleted central cavity, indicative of significant flux deficit at the centre of the disc. Ring discs, in contrast, exhibit alternating patterns of dust enhancements and depletions that create distinct annular ring structures, without a central cavity. Highly-inclined discs  (>75$^{\circ}$ inclination), where their orientation conceals internal features, are classified as inclined discs, emphasising the impact of viewing geometry on their appearance. Discs appearing more face-on that lack discernible substructures and appear featureless in the continuum images are classified as discs lacking visual substructures. \\
\indent Based on these visual classifications, we identified five transition discs, four ring discs, four inclined discs, and twelve discs lacking visual substructure. One disc, 2MASS J11095340-7634255, could not be categorised due to its complex morphology, featuring a complex object adjacent to an extended, spiral-like structure. Based on the current position of the central star, indicated by a white star in the cleaned image, it is likely that the more extended object represents the protoplanetary disc surrounding the central star. However, the continuum image alone does not provide sufficient information to identify the nature of the adjacent compact object.

\section{Modelling disc substructures}
\label{sec:modelling}
\subsection{Modelling approach}
\label{subsec:modelling_approach}
The dust continuum data were analysed in the visibility plane to assess the presence of substructures in all protoplanetary discs in our sample. Each disc was modelled as a sum of axisymmetric radial Gaussian rings, with the intensity profile expressed as

\begin{equation}
\hspace*{0.25\columnwidth}
    I(r) = \sum_{i=1}^{n} I_{i} \exp\left[-\left(\frac{r - r_{i}}{w_{i}}\right)^{2}\right],
\label{eq:gaussian}
\end{equation}

\noindent where \textit{I$_{i}$} represents the peak intensity of the \textit{i}th substructure, \textit{r} is the radial distance from the disc centre, \textit{r$_{i}$} denotes the location of the substructure, and \textit{w$_{i}$} is the width of the Gaussian curve, where the substructure width is defind by the full width at half maximum (FWHM) of the profile. The corresponding model visibilities were obtained by Fourier-transforming the intensity profile using \texttt{Galario} (\citealt{Galario}). The binary discs were modelled using the \texttt{sampleImage} rather than the \texttt{sampleProfile} function in \texttt{Galario} in order to model both discs simultaneously.\\
\indent Model fitting was performed in the visibility domain using \texttt{emcee} (\citealt{emcee}), a Markov Chain Monte Carlo (MCMC) sampler that explored the optimal values of the free parameters. For each disc, the four geometric parameters (disc inclination angle (\textit{i}), disc position angle (PA), and position offsets from the phase centre ($\Delta$RA and $\Delta$Dec)) and the parameters of the Gaussian profiles were fit. To determine the number, locations and widths of the rings, we employed \texttt{Frankenstein} (\citealt{Frankenstein}), which reconstructs radial brightness profiles directly from the visibilities. The disc geometry and radial intensity profiles derived from \texttt{Frankenstein} were used as initial guesses for the \texttt{Galario} models, which were subsequently refined via MCMC.\\
\indent On average, the MCMC fitting used 60 random walkers, each running for 3000 iterations. Initial parameter values were perturbed using a normal distribution scaled by a factor of 10$^{\text{-4}}$ to ensure broad exploration of the parameter space. The burn-in phase typically required 2000 steps, after which the posterior distributions were sampled over the final 1000 iterations. The best-fit parameters were taken as the median values of the posterior distributions, with uncertainties estimated from the 16th-84th percentile range.\\
\indent Due to its complex morphology, 2MASS J11095340-7634255  could not be reliably modelled.

\subsection{Modelling results}
\label{subsec:modelling_results}
\begin{figure*}[hbt]
%\hspace{-5mm}
    %\centering
    \includegraphics[scale=0.19]{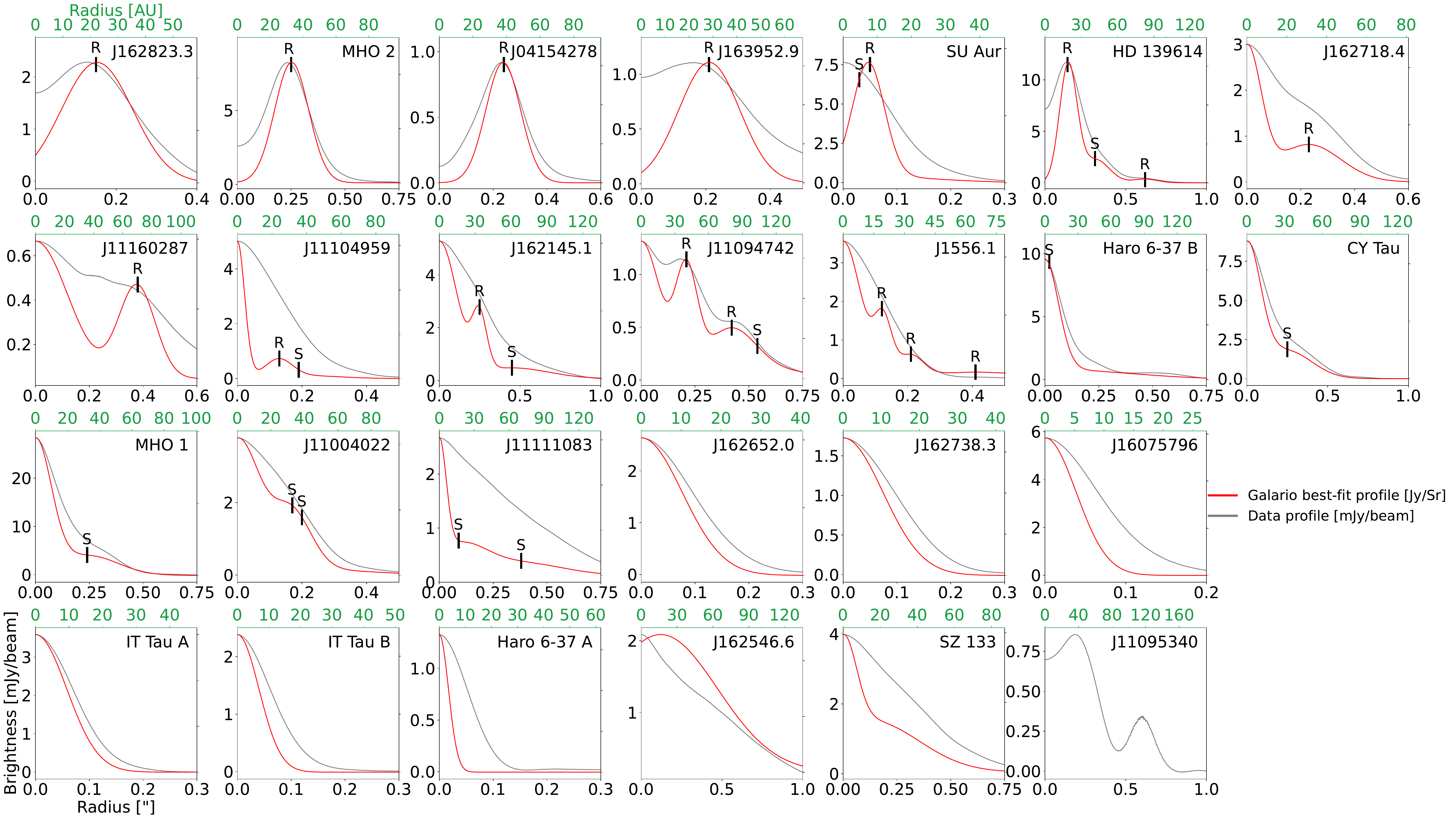}
    \caption{Azimuthally averaged radial brightness profiles for all 26 discs (grey), overlaid with the best-fit model profiles (red) for the 25 modelled discs. Ring and shoulder locations in the models are marked with \textbf{R} and \textbf{S}, respectively. The corresponding physical radii in AU are indicated in green above each panel. For discs with names starting with SSTc2d, 2MASS, or RX, the names in the upper right corners are abbreviated to the first part after "J" (e.g., SSTc2d J162823.3-242241 is labelled as J162823.3).}
    \label{fig:radial_profiles}
\end{figure*}

\begin{table*}[hbt]
    \caption{Best-fit disc morphologies and dust disc radii for all studied discs. Disc morphologies and geometries are derived using \texttt{Galario}, while the effective dust radii ($R_{68\%}$ and $R_{95\%}$) are derived from the curve-of-growth method. Column 1 lists the source names. Columns 2 and 3 provide the 68$\%$ effective dust radii from this study and previously reported values (upper limits denoted by "<"). Column 4 gives the 95$\%$ effective dust radii. Columns 5 and 6 show the MCMC-fitted inclination and position angle. Column 7 describes the best-fit disc morphology.}
    \hspace{-3mm}
    \begin{tabular}{l c c c c c l}
        \hline\hline
        \textbf{Disc Name} & \multicolumn{2}{c}{\textbf{\textit{R$_{\textbf{68\%}}$}}} &\textbf{\textit{R$_{\textbf{95\%}}$}} & \textbf{\textit{i}} & \textbf{PA} & \textbf{Morphology/Model Description}\\
        &  & \hspace{-20mm}\textbf{[AU]}& \textbf{[AU]} & \textbf{[$^{\circ}$]} & \textbf{[$^{\circ}$]} &\\
        & This Study & Previous & & & &\\
        
        \hline
        SSTc2d J162823.3-242241 & 60 & 62 (a) & 92 & 68.46 & 66.23 & Inner cavity \\
        MHO 2 & 49 & 37 (c) & 66 & 45.54 & 106.25 & Inner cavity \\
        2MASS J04154278+2909597 & 50 & 39 (b) & 69 & 33.21 & 174.94 & Inner cavity \\
        SSTc2d J163952.9-241931 & 43 & 31 (a) & 58 & 81.41 & 106.31 & Inner cavity \\
        SU Aur & 23 & 50 (c) & 42 & 42.64 & 117.88 & Inner cavity + one shoulder \\
        HD 139614 & 44 & 67 (d) & 87 & 0.46 & -14.91 & Inner cavity + one ring + one shoulder \\
        SSTc2d J162718.4-243915 & 52 & 35 (a) & 76 & 69.93 & 53.88 & Inner Gaussian profile + one ring \\
        2MASS J11160287-7624533 & 81 & 39 (e) & 121 & 82.29 & 133.57 & Inner Gaussian profile + one ring \\
        2MASS J11104959-7717517 & 48 & 31 (e) & 88 & 49.28 & 62.32 & Inner Gaussian profile + one ring + one shoulder \\
        SSTc2d J162145.1-234232 & 81 & 36 (a) & 124 & 72.88 & 174.93 & Inner Gaussian profile + one ring + one shoulder \\
        2MASS J11094742-7726290 & 92 & 72 (e) & 140 & 51.43 & 27.82 & Inner Gaussian profile + two rings + one shoulder \\
        RX J1556.1-3655 & 33 & 42 (f) & 57 & 49.79 & 56.74 & Inner Gaussian profile + three rings \\
        Haro 6-37 B & 98 & 34 (c) & 150 & 27.95 & 179.98 & Inner Gaussian profile + one shoulder\\
        CY Tau & 47 & 48 (b) & 75 & 15.34 & 160.21 & Inner Gaussian profile + one shoulder \\
        MHO 1 & 48 & 26 (c) & 78 & 39.40 & 83.98 & Inner Gaussian profile + one shoulder \\
        2MASS J11004022-7619280 & 48 & 36 (e) & 89 & 16.74 & -3.23 & Inner Gaussian profile + two shoulders \\
        2MASS J11111083-7641574 & 98 & 68 (e) & 140 & 79.49 & 141.81 & Inner Gaussian profile + two shoulders \\
        SSTc2d J162652.0-243039 & 21 & 51 (a) & 38 & 39.31 & 96.32 & Compact \\
        SSTc2d J162738.3-235732 & 21 & <14 (a) & 34 & 38.63 & 92.78 & Compact \\
        2MASS J16075796-2040087 & 16 & <34 (g) & 29 & 56.55 & 10.48 & Compact \\
        IT Tau A & 21 & <32 (c) & 30 & 63.80 & 51.24 & Compact \\
        IT Tau B & 21 & <32 (c) & 36 & 51.89 & 62.63 & Compact \\
        Haro 6-37 A & 17 & 13 (c) & 23 & 57.38 & -4.48 & Compact \\
        SSTc2d J162546.6-242336 & 89 & 48 (a) & 128 & 80.88 & 112.90 & Highly-inclined \\
        SZ 133 & 59 & 104 (f) & 87 & 77.18 & 126.10 & Highly-inclined \\
        \hline
    \end{tabular}
    \caption*{\textbf{References.} (a) \citet{Cieza}, (b) \citet{tripathi}, (c) \citet{Akeson}, (d) \citet{Vioque}, (e) \citet{Pascutti}, (f) \citet{Ansdell}, (g) \citet{Barenfeld}}
\label{tab:galario}
\end{table*}

\begin{table*}[hbt]
    \caption{Best-fit ring and shoulder properties for discs with substructures. Column 1 lists the source names, and column 2 specifies the substructure type. Columns 3 and 4 provide the central location in arcseconds and AU, respectively, while columns 5 and 6 give the full width at half maximum in the same units.}
    \centering
    \renewcommand{\arraystretch}{1.2}
    \begin{tabular}{l c c c c c}
        \hline\hline
        \vspace{-2mm}
        \textbf{Disc Name} &\textbf{Substructure} & \multicolumn{2}{c}{\textbf{Location}} & \multicolumn{2}{c}{\textbf{Width}}\\

        & & \textbf{["]} & \textbf{[AU]} & \textbf{["]} & \textbf{[AU]}\\
        \hline
        SSTc2d J162823.3-242241 & Ring &  0.15$\small\substack{\text{+0.00} \\ \text{-0.00}}$ &  22.05$\small\substack{\text{+0.00} \\ \text{-0.00}}$ &  0.22$\small\substack{\text{+0.00} \\ \text{-0.00}}$ & 31.78$\small\substack{\text{+0.00} \\ \text{-0.00}}$\\
        MHO 2 & Ring &  0.25$\small\substack{\text{+0.00} \\ \text{-0.00}}$ &  33.02$\small\substack{\text{+0.39} \\ \text{-0.13}}$ &  0.08$\small\substack{\text{+0.00} \\ \text{-0.00}}$ & 10.35$\small\substack{\text{+0.13} \\ \text{-0.13}}$\\
        2MASS J04154278+2909597 & Ring & 0.24$\small\substack{\text{+0.00} \\ \text{-0.00}}$ &  38.42$\small\substack{\text{+0.00} \\ \text{-0.00}}$ &  0.15$\small\substack{\text{+0.00} \\ \text{-0.00}}$ & 23.99$\small\substack{\text{+0.00} \\ \text{-0.00}}$\\
        SSTc2d J163952.9-241931 & Ring & 0.21$\small\substack{\text{+0.00} \\ \text{-0.00}}$ &  28.35$\small\substack{\text{+0.00} \\ \text{-0.00}}$ &  0.22$\small\substack{\text{+0.02} \\ \text{-0.02}}$ & 29.22$\small\substack{\text{+2.25} \\ \text{-2.25}}$\\
        SU Aur & Shoulder & 0.03$\small\substack{\text{+0.04} \\ \text{-0.02}}$ &  4.71$\small\substack{\text{+6.28} \\ \text{-3.14}}$ &  0.30$\small\substack{\text{+0.02} \\ \text{-0.03}}$ & 47.05$\small\substack{\text{+2.61} \\ \text{-5.23}}$\\
        & Ring & 0.05$\small\substack{\text{+0.00} \\ \text{-0.00}}$ &  7.85$\small\substack{\text{+0.00} \\ \text{-0.00}}$ &  0.07$\small\substack{\text{+0.00} \\ \text{-0.00}}$ & 10.46$\small\substack{\text{+0.00} \\ \text{-0.00}}$\\
        HD 139614 & Ring & 0.14$\small\substack{\text{+0.00} \\ \text{-0.00}}$ &  18.71$\small\substack{\text{+0.00} \\ \text{-0.00}}$ &  0.13$\small\substack{\text{+0.00} \\ \text{-0.00}}$ & 17.80$\small\substack{\text{+0.00} \\ \text{-0.00}}$\\
        & Shoulder & 0.31$\small\substack{\text{+0.00} \\ \text{-0.00}}$ &  41.42$\small\substack{\text{+0.00} \\ \text{-0.00}}$ &  0.12$\small\substack{\text{+0.00} \\ \text{-0.00}}$ & 16.03$\small\substack{\text{+0.00} \\ \text{-0.00}}$\\
        & Ring & 0.62$\small\substack{\text{+0.00} \\ \text{-0.00}}$ &  82.84$\small\substack{\text{+0.00} \\ \text{-0.00}}$ &  0.18$\small\substack{\text{+0.02} \\ \text{-0.00}}$ & 24.47$\small\substack{\text{+2.22} \\ \text{-0.00}}$\\
        SSTc2d J162718.4-243915 & Ring &  0.23$\small\substack{\text{+0.01} \\ \text{-0.01}}$ &  31.05$\small\substack{\text{+1.35} \\ \text{-1.35}}$ &  0.28$\small\substack{\text{+0.00} \\ \text{-0.00}}$ & 38.21$\small\substack{\text{+0.00} \\ \text{-0.00}}$\\
        2MASS J11160287-7624533 & Ring &  0.38$\small\substack{\text{+0.01} \\ \text{-0.01}}$ &  70.30$\small\substack{\text{+1.85} \\ \text{-1.85}}$ &  0.17$\small\substack{\text{+0.02} \\ \text{-0.02}}$ & 30.80$\small\substack{\text{+3.08} \\ \text{-3.08}}$\\
        2MASS J11104959-7717517 & Ring &  0.13$\small\substack{\text{+0.01} \\ \text{-0.01}}$ &  24.27$\small\substack{\text{+1.87} \\ \text{-1.87}}$ &  0.10$\small\substack{\text{+0.02} \\ \text{-0.02}}$ & 18.65$\small\substack{\text{+3.11} \\ \text{-3.11}}$\\
        & Shoulder &  0.19$\small\substack{\text{+0.06} \\ \text{-0.08}}$ &  35.47$\small\substack{\text{+11.20} \\ \text{-14.94}}$ &  0.30$\small\substack{\text{+0.07} \\ \text{-0.05}}$ & 55.95$\small\substack{\text{+12.43} \\ \text{-9.33}}$\\
        SSTc2d J162145.1-234232 & Ring &  0.25$\small\substack{\text{+0.00} \\ \text{-0.01}}$ &  33.75$\small\substack{\text{+0.00} \\ \text{-1.35}}$ &  0.10$\small\substack{\text{+0.02} \\ \text{-0.02}}$ & 13.49$\small\substack{\text{+2.25} \\ \text{-2.25}}$\\
        & Shoulder &  0.45$\small\substack{\text{+0.02} \\ \text{-0.02}}$ &  60.75$\small\substack{\text{+2.70} \\ \text{-2.70}}$ &  0.53$\small\substack{\text{+0.03} \\ \text{-0.03}}$ & 71.93$\small\substack{\text{+4.50} \\ \text{-4.50}}$\\
        2MASS J11094742-7726290 & Ring &  0.21$\small\substack{\text{+0.00} \\ \text{-0.01}}$ &  40.20$\small\substack{\text{+0.00} \\ \text{-1.91}}$ &  0.12$\small\substack{\text{+0.02} \\ \text{-0.02}}$ & 22.31$\small\substack{\text{+3.19} \\ \text{-3.19}}$\\
        & Ring &  0.42$\small\substack{\text{+0.01} \\ \text{-0.01}}$ &  80.40$\small\substack{\text{+1.91} \\ \text{-1.91}}$ &  0.27$\small\substack{\text{+0.02} \\ \text{-0.02}}$ & 51.00$\small\substack{\text{+3.19} \\ \text{-3.19}}$\\
        & Shoulder &  0.54$\small\substack{\text{+0.06} \\ \text{-0.05}}$ &  103.37$\small\substack{\text{+11.49} \\ \text{-9.57}}$ &  0.57$\small\substack{\text{+0.07} \\ \text{-0.05}}$ & 108.37$\small\substack{\text{+12.75} \\ \text{-9.56}}$\\
        RX J1556.1-3655 & Ring &  0.12$\small\substack{\text{+0.01} \\ \text{-0.01}}$ &  18.95$\small\substack{\text{+1.58} \\ \text{-1.58}}$ &  0.07$\small\substack{\text{+0.03} \\ \text{-0.02}}$ & 10.52$\small\substack{\text{+5.26} \\ \text{-2.63}}$\\
        & Ring &  0.21$\small\substack{\text{+0.02} \\ \text{-0.03}}$ &  33.17$\small\substack{\text{+3.16} \\ \text{-4.74}}$ &  0.08$\small\substack{\text{+0.03} \\ \text{-0.03}}$ & 13.15$\small\substack{\text{+5.26} \\ \text{-5.26}}$\\
        & Ring &  0.41$\small\substack{\text{+0.01} \\ \text{-0.02}}$ &  64.76$\small\substack{\text{+1.58} \\ \text{-3.16}}$ &  0.10$\small\substack{\text{+0.07} \\ \text{-0.05}}$ & 15.78$\small\substack{\text{+10.52} \\ \text{-7.89}}$\\
        Haro 6-37 B & Shoulder &  0.01$\small\substack{\text{+0.00} \\ \text{-0.00}}$ &  0.20$\small\substack{\text{+0.20} \\ \text{-0.20}}$ &  0.38$\small\substack{\text{+0.00} \\ \text{-0.00}}$ & 73.48$\small\substack{\text{+0.20} \\ \text{-0.20}}$\\
        CY Tau & Shoulder &  0.25$\small\substack{\text{+0.00} \\ \text{-0.00}}$ &  32.10$\small\substack{\text{+0.00} \\ \text{-0.00}}$ &  0.20$\small\substack{\text{+0.00} \\ \text{-0.00}}$ & 25.66$\small\substack{\text{+0.00} \\ \text{-0.00}}$\\
        MHO 1 & Shoulder &  0.24$\small\substack{\text{+0.00} \\ \text{-0.00}}$ &  2.55$\small\substack{\text{+0.54} \\ \text{-0.54}}$ &  0.14$\small\substack{\text{+0.00} \\ \text{-0.00}}$ & 18.89$\small\substack{\text{+0.27} \\ \text{-0.27}}$\\
        2MASS J11004022-7619280 & Shoulder &  0.17$\small\substack{\text{+0.02} \\ \text{-0.02}}$ &  32.81$\small\substack{\text{+3.86} \\ \text{-3.86}}$ &  0.15$\small\substack{\text{+0.02} \\ \text{-0.02}}$ & 28.92$\small\substack{\text{+3.21} \\ \text{-3.21}}$\\
        & Shoulder &  0.20$\small\substack{\text{+0.04} \\ \text{-0.03}}$ &  38.60$\small\substack{\text{+7.72} \\ \text{-5.79}}$ &  0.48$\small\substack{\text{+0.03} \\ \text{-0.03}}$ & 93.20$\small\substack{\text{+6.43} \\ \text{-6.43}}$\\
        2MASS J11111083-7641574 & Shoulder &  0.09$\small\substack{\text{+0.05} \\ \text{-0.05}}$ &  16.65$\small\substack{\text{+9.25} \\ \text{-9.25}}$ &  0.28$\small\substack{\text{+0.10} \\ \text{-0.08}}$ & 52.37$\small\substack{\text{+18.48} \\ \text{-15.40}}$\\
        & Shoulder &  0.38$\small\substack{\text{+0.06} \\ \text{-0.05}}$ &  70.30$\small\substack{\text{+11.10} \\ \text{-9.25}}$ &  0.45$\small\substack{\text{+0.05} \\ \text{-0.05}}$ & 83.17$\small\substack{\text{+9.24} \\ \text{-9.24}}$\\
        
        \hline
    \end{tabular}
\label{tab:loc_width}
\end{table*}
We selected the best-fit model for each disc by using the fewest parameters necessary to reproduce axisymmetric dust features with residuals below $\sim$3$\sigma$. Figures \ref{fig:galario_first}, \ref{fig:galario_second}, and \ref{fig:galario_third} (Appendix \ref{appendix:Best-fit models}) compare the best-fit models with the observed visibilities and cleaned images, along with the corresponding radial brightness profiles. Figure \ref{fig:radial_profiles} presents the best-fit radial brightness profiles for the modelled 25 discs, along with the azimuthally averaged profiles for all 26 discs, while Table \ref{tab:galario} summarises the best-fit morphologies, effective dust disc radii, incinations, and position angles. Substructure locations and widths are detailed in Table \ref{tab:loc_width} and are highlighted in Figure \ref{fig:radial_profiles}.

\subsection{Best-fit models and residuals}
Overall, the models reproduce the observations well, with minimal residuals. Some discs (e.g. SU Aur and RX J1556.1-3655) exhibit slight azimuthal asymmetries, likely caused by shadows from warps or a misaligned inner disc. Such warps can result from planetary interactions (\citealt{Zhang2018}) or internal instabilities (\citealt{Kratter}), while inner-disc misalignments may arise from tilted stellar magnetic fields (\citealt{Bouvier}). Notably, HD 139614 leaves a crescent-like residual, suggesting an unresolved azimuthal asymmetry not captured by simple radial Gaussian rings. Haro 6-37 B, MHO 1, and MHO 2 also exhibit more substantial residuals, indicating that binary discs are more difficult to model accurately and may harbour additional substructures not resolved in our current axisymmetric framework.\\
\indent For SSTc2d J162546.6-242336 and SZ 133, the models fail to accurately reproduce the data due to their high inclinations ($\sim$81$^{\circ}$ and $\sim$77$^{\circ}$, see Table \ref{tab:galario}). At this angular resolution, resolving ring features in highly-inclined discs proved challenging, preventing model convergence. Consequently, their best-fit morphologies do not fully reflect their true dust distributions.

\subsection{Dust disc sizes}
The effective dust disc radii in Table \ref{tab:galario} correspond to the radii encircling 68$\%$ ($R_{68\%}$) and 95$\%$ ($R_{95\%}$) of the total flux density. In most cases, our measured disc radii are smaller or larger than previously reported values, likely due to differences in angular resolution, uv-coverage, or sensitivity between observations. Higher-resolution data can resolve more compact structures and reveal substructures that alter flux distribution, while lower-resolution data may smooth out features and overestimate disc size. Variations in disc inclination, orientation, and image reconstruction techniques may also contribute to discrepancies.\\
\indent For more inclined discs, the measured radius may be larger due to improved inclination constraints. Low-resolution data can misinterpret an edge-on disc as more face-on, leading to an underestimation of its size. Our high-resolution measurements provide a more accurate representation of true disc sizes by resolving inclination effects.

\subsection{Disc classifications}
We classify the modelled discs into five categories: transition (TD), ring (RD), shoulder (SD), compact (CD), and highly-inclined discs (HID), as detailed below. The final disc classifications of all 26 discs can be found in column 7 of Table \ref{tab:cleaned}.%Transition discs feature a central cavity with an offset peak in their radial brightness profiles. Ring discs exhibit one or more rings with distinct peaks and dips, while shoulder discs lack clear minima and maxima, resembling smooth transitions. Compact discs are best described by a single Gaussian profile and highly-inclined discs are those that could not be accurately modelled due to their nearly edge-on view. Based on these criteria, we identify five transition, six ring, three shoulder, three compact, and two highly-inclined discs.

\subsubsection{Transition discs}
The transition discs %(SSTc2d J162823.3-242241, 2MASS J04154278+2909597, SSTc2d J163952.9-241931, SU Aur, and HD 139614)
all exhibit central cavities with surrounding rings at varying radii. However, while the best-fit model of SU Aur suggests a cavity, the observed image does not clearly show one, though a dip in the visibility profile supports its presence. HD 139614 and SU Aur also feature shoulders, suggesting additional substructures. HD 139614 has previously been identified as a pre-transitional disc based on its spectral energy distribution (SED), which shows a mid-infrared dip indicative of a dust-depleted inner region (\citealt{Labadie2014}). Similarly, SU Aur's SED has been modelled with a partially cleared inner disc, consistent with the presence of a cavity inferred from our visibility profile analysis (\citealt{Akeson2002}; 
\citealt{Jeffers2014}). Finally, the transition disc cavity 2MASS J04154278+2909597 was previously recovered with the SMA (Submillimeter Array) (\citealt{Espaillat2015}) and recently, a misaligned transiting planet was discovered in this system, although at an orbit that is way too small to explain the cavity size (\citealt{Barber2024}). The misaligned architecture of the planet-disc system has been analysed in more detail in other works (\citealt{Huhn}; \citealt{Shoshi}).

%\subsubsection{Ring discs}
%The ring discs are SSTc2d J162718.4-243915, 2MASS J11160287-7624533, 2MASS J11104959-7717517, SSTc2d J162145.1-234232, 2MASS J11094742-7726290, and RX J1556.1-3655. Ring discs contain one to three rings at different radii, with some also exhibiting shoulders.

\subsubsection{Ring and shoulder discs}
Ring discs contain one to three rings at different radii, with some also exhibiting shoulders. The shoulder discs %are CY Tau, 2MASS J11004022-7619280, and 2MASS J11111083-7641574 and 
display smooth brightness transitions rather than distinct rings. These inflection points (\citealt{Cieza2021}) could result from unresolved gaps, weak pressure bumps (\citealt{Zhu}; \citealt{Sturm}), or grain size variations, potentially linked to the presence of snowlines (\citealt{Cieza2016}). The presence of shoulders alongside rings, gaps, and cavities in other discs suggests that they may represent an intermediate stage in disc evolution. Rather than being static structures, shoulders may arise from gradual dust accumulation at pressure maxima that are not yet deep or sharp enough to open a gap or form a high-contrast ring.

\subsubsection{Compact discs}
%SSTc2d J162652.0-243039, SSTc2d J162738.3-235732, and 2MASS J16075796-2040087 are 
Compact discs in this subsection are defined as discs with an effective dust radius $R_{\mathrm{68\%,dust}}<30$ AU that do not exhibit detectable substructures at the current angular resolution. These discs are modelled by a single Gaussian profile. Higher-resolution observations may reveal narrow substructures or substructures with low contrast.

\subsubsection{Highly-inclined discs}
SSTc2d J162546.6-242336 and SZ 133 could not be accurately modelled due to their near edge-on orientation. Although five discs in our sample have inclinations >75$^{\circ}$ (see Table \ref{tab:galario}), three were successfully modelled, likely due to the strong contrast between their substructures and their surrounding material. The two remaining discs may contain shallow gaps or faint rings that are difficult to discern due to their nearly edge-on orientations. Despite these challenges, edge-on discs provide valuable insights into dust dynamics. \citet{Villenave2020} demonstrated that vertical dust settling and radial drift can be observed in such systems, with shorter wavelengths tracing larger disc sizes, potentially indicating dust traps. A similar analysis of our highly-inclined discs could reveal whether dust traps are present.

\section{Discussion}
\label{sec:Discussion}
\subsection{Substructures in this sample}
Of the 26 protoplanetary discs in our sample, the cleaned images and best-fit \texttt{Galario} models reveal a diverse range of morphologies, seventeen discs exhibit substructures: six are transition discs, six are ring discs, and five are shoulder discs. The remaining nine lack clear substructures at the current resolution: six have $R_{\mathrm{68\%,dust}}<30$ AU and are classified as compact, two are highly inclined and may obscure substructures, and one (2MASS J11095340-7634255) has an unexplained dust morphology.
%Further research is needed to determine whether MHO 1 hosts substructures. If it is indeed smooth, the fact that it is part of a binary system could offer a potential explanation. Binary interactions are known to truncate discs, enhance radial drift, and shorten disc lifetimes, which may reduce the window for substructure formation or shift to smaller scales that remain unresolved (e.g. \citealt{Manara2019}; \citealt{Zagaria2023}).\\
%\indent However, substructures are not absent in all binary discs. In our sample, MHO 2 exhibits a cavity, and Haro 6-37 B shows a clear ring structure. This indicates that substructures can still emerge in binary systems \textbf{as seen in previous works \citep[e.g.][]{Andrews2018,Menard2020,Kurtovic2024}}, possibly triggered by resonances, spiral density waves, or planets forming in dynamically stable regions despite the complex conditions introduced by a stellar companion (e.g \citealt{Picogna}; \citealt{Silsbee}).
\subsection{The radial distribution of substructures}
Discs with substructures in our sample have 68$\%$ effective dust radii ranging from 23-98 AU and 95$\%$ radii from 42-150 AU. In contrast, all compact discs without detected substructures have 68$\%$ radii smaller than 22 AU and 95$\%$ radii $\leq$38 AU. Apart from SSTc2d J162546.6-242336, SZ 133, and 2MASS J11095340-7634255, where the presence of substructures remains uncertain, every disc with 68$\%$ and 95$\%$ radii exceeding 21 and 38 AU, respectively, exhibit rings, gaps, or shoulders.\\
\indent These findings align with previous ALMA disc surveys at similar resolutions ($\sim$0.12"). The Taurus ALMA survey (\citealt{Long, Long2019}) found that discs with substructures typically have 95$\%$ dust radii between $\sim$40 and $\sim$200 AU, while the DSHARP survey (\citealt{Huang}) reported substructured discs with 95$\%$ radii spanning $\sim$30 to $\sim$200 AU. This consistency suggests that large-scale substructures are commonly detected in extended dust discs when observed at sufficient angular resolution across multiple star-forming regions.\\ %a strong correlation between disc size and the presence of substructures across multiple star-forming regions.\\
\begin{figure}[hbt]
    \includegraphics[scale=0.36]{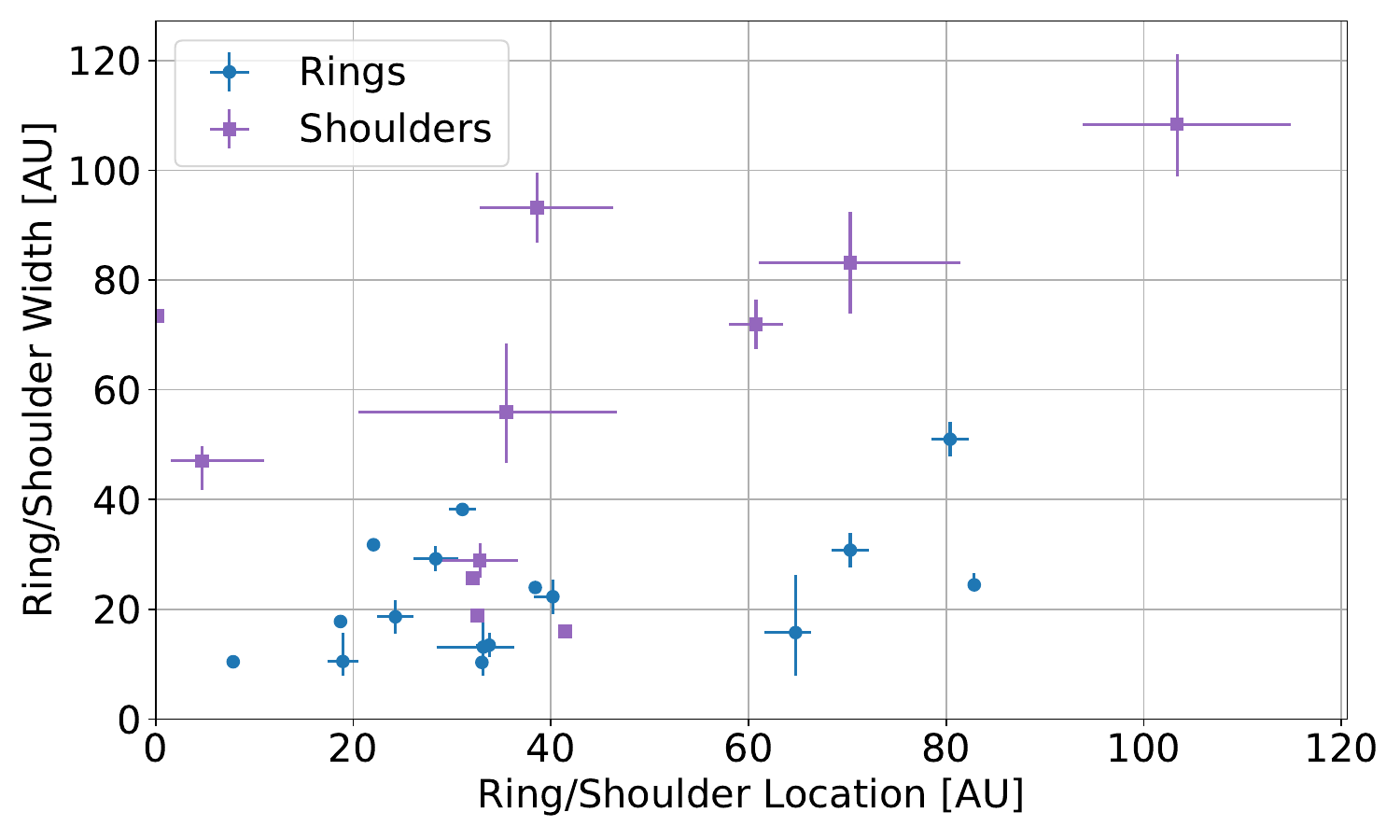}
    \caption{Ring and shoulder widths as a function of their radial locations. Ring substructures are shown in blue, and shoulder substructures are shown in purple.}
    \label{fig:location_vs_width}
\end{figure}
\indent Figure \ref{fig:location_vs_width} shows the distribution of ring and shoulder positions in our sample, which span a wide radial range from $\sim$0.2 to $\sim$103 AU. The widths of the substructures in our sample vary from $\sim$10 to $\sim$108 AU, indicating a broad diversity in morphology. Substructures are found across nearly the full radial extent of the discs, suggesting that ring and shoulder formation can occur under a wide range of physical conditions, without a strong preference for specific radii similar as seen in previous work \citep[e.g.][]{Andrews2018,Long,vandermarel2019}.

\subsection{2MASS J11095340-7634255}
\begin{figure}[hbt]
    \includegraphics[scale=0.25]{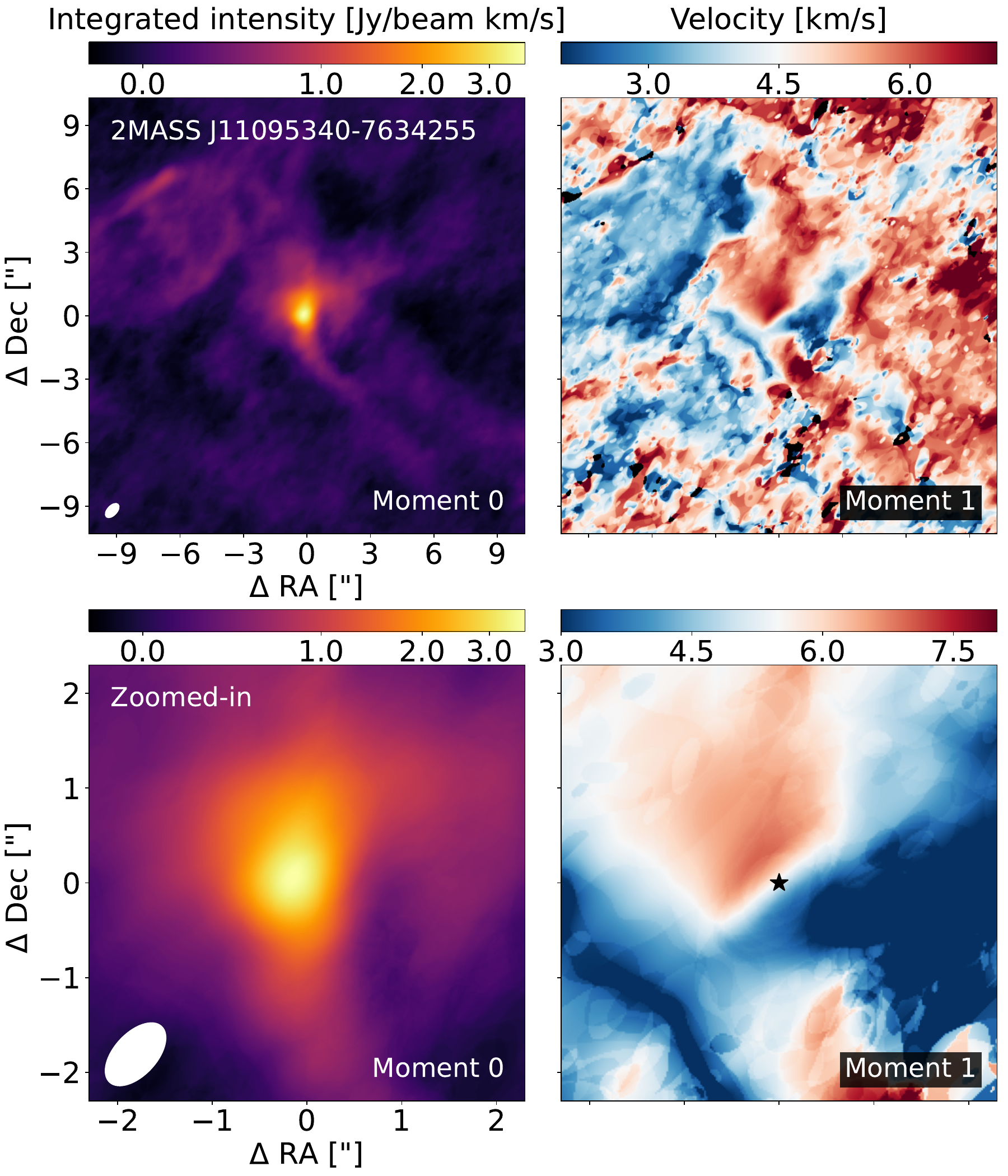}
    \caption{Gallery of the high-resolution $^{12}$CO J=2-1 moment maps for 2MASS J11095340-7634255, constructed from emission exceeding the 3$\sigma$ threshold. The top row shows a $20.0\text{"}\times20.0\text{"}$ field of view, while the bottom row provides a zoomed-in view of a $4.6\text{"}\times4.6\text{"}$ region around the disc. The moment 0 maps (left column) display integrated intensity in Jy/beam km/s, and the moment 1 maps (right column) show velocity in km/s. Axes indicate RA and Dec offsets from the disc centre in arcseconds. The white ellipse in the moment 0 maps represents the beam size. The black star in the zoomed-in moment 1 map marks the current position of the central star.}
    \label{fig:CO_2mass_255}
\end{figure}

Figure \ref{fig:CO_2mass_255} presents the $^{12}$CO moment 0 and moment 1 maps for 2MASS J11095340-7634255, the disc with an unexplained dust continuum morphology, observed at low angular resolution, with the bottom row offering a zoomed-in view. The moment 0 map reveals significant CO emission surrounding the star, with bright, concentrated emission in the central region of the disc. In the moment 1 map, various non-Keplerian velocity patterns suggest interactions between the disc and the surrounding gas. The zoomed-in moment 1 map highlights a velocity gradient around the marked position of the central star, where a dividing line separates redshifted and blueshifted emission, confirming Keplerian rotation within the disc.\\
\indent The continuum and gas morphology both indicate that the protoplanetary disc is likely interacting with a nearby compact structure. Although no second stellar source is detected, the morphology is reminiscent of close binaries. We propose that 2MASS J11095340-7634255 could potentially be a super-imposed binary system, meaning two closely projected protoplanetary discs, with the central stars nearly aligned along the line of sight, causing one to remain undetected, while the discs themselves are offset and closely interacting. Alternatively, it could be a sign of interaction with ambient material, in particular late-stage infall \citep{Kuffmeier2023}. %If confirmed, this configuration would have implications for our understanding of early disc evolution in crowded environments, where disc-disc interactions can shape morphology and dynamics. 
The observed multi-wavelength variability and asymmetric gas emission (\citealt{Feigelson}, \citealt{RodgersLee}, \citealt{Sturm}) further support the presence of ongoing dynamical processes within the system. Given these characteristics and the apparent extended, non-axisymmetric morphology, this source is provisionally classified as an extended disc (ED in Table \ref{tab:cleaned}). Further observations are needed to assess the presence of a companion and to clarify the system’s structure, allowing us to distinguish between these two scenarios.

\subsection{Complete sample of extended protoplanetary discs}
\begin{table*}[hbt]
    \centering
    \caption{Updated classification on discs >30 AU from \citet{NienkeGijs}. Column 1 lists the source names, and column 2 the  star-forming regions (with "Upper Sco" representing Upper Scorpius). Column 3 provides the 68\% effective dust disc radii (AU), with upper limits denoted by "<". The disc classifications are listed in column 4: TD (transition disc), RD (ring disc), SD (shoulder disc), HID (highly-inclined disc), CD (compact disc), and ED (extended disc), and column 5 includes the references for all listed values.}
    \begin{tabular}{lcccl}
\hline \hline   
    \textbf{Disc Name} & \textbf{Region} & \textbf{\textit{R$_{\textbf{68\%}}$}} & \textbf{Class} & \textbf{Reference} \\
    &&\textbf{[AU]}&&\\
    \hline
2MASS J10590108-7722407 & Chamaeleon	&		37 &	TD	&	ALMA program 2021.1.00854.S	\\
2MASS J11074245-7733593 & Chamaeleon	&		<31 &	CD (unresolved)	&	\citet{Pascucci2016}	\\
2MASS J11432669-7804454 & Chamaeleon	&		- &	Non-detection	&	\citet{Pascucci2016}, ALMA program 2023.1.01276.S	\\
EX Lup & Lupus	&	62	&	RD	&	\citet{guerra2025}	\\
SSTc2d J160002.4-422216 & Lupus	&	86	&	HID	&	\citet{guerra2025}	\\
Sz 65 & Lupus	&	27	&	CD	&	\citet{Miley2024}	\\
Sz 73 &  Lupus	&	42	&	RD	&	\citet{guerra2025}	\\
Sz 98 & Lupus	&	106	&	RD	&	\citet{Gasman2023}	\\
Sz 76 & Lupus	&	16	&	TD	&	\citet{guerra2025}	\\
SSTc2d J162623.6-242439 & Ophiuchus	&	91	&	HID	&	\citet{Cieza2021}	\\
SSTc2d J162640.5-242714 & Ophiuchus	&	119	&	RD	&	\citet{Cieza2021}	\\
SSTc2d J163131.2-242628 & Ophiuchus	&	150	&	HID	&	\citet{Villenave2022}	\\
SSTc2d J163135.6-240129 & Ophiuchus	&	67	&	RD	&	\citet{Cieza2021}	\\
DR Tau & Taurus	&	51	&	SD	&	\citet{Yamaguchi2024}	\\
2MASS J04141700+2810578 & Taurus	& 44		&	RD	&	\citet{Yamaguchi2024}	\\
2MASS J04202144+2813491 & Taurus	&	263	&	HID	&	\citet{Villenave2020}	\\ %Band 6 image: 3.76" major axis size, 140 pc assumed: 263 au!
J04230776+2805573 & Taurus	&	70	&	HID	&	\citet{Villenave2020}	\\ %B7, 1.0" major axis size, 140 pc: 
2MASS J04322210+1827426 & Taurus	&	34	&	TD	&	\citet{Kurtovic2021}	\\ %mho6
2MASS J04334465+2615005 & Taurus	&	37	&	RD	&	\citet{Kurtovic2021}	\\
V409 Tau & Taurus	&	42	&	SD	&	\citet{Yamaguchi2024}	\\
V710 Tau A & Taurus	&	43	&	SD	&	\citet{Yamaguchi2024}	\\
2MASS J04265440+2606510 & Taurus	&	<39	&	CD (unresolved)	&	\citet{Akeson2019}	\\
2MASS J04295950+2433078 & Taurus	&	<17	&	CD (unresolved)	&	ALMA program 2016.1.01511.S	\\
2MASS J04324911+2253027 & Taurus	&	<18	&	CD (unresolved)	&	ALMA program 2016.1.01511.S	\\
AS 205 N & Upper Sco	&	35	&	ED	&	\citet{Andrews2018}	\\
\hline
    \end{tabular}
    \label{tab:update}
\end{table*}

With the new analysis of 26 discs at high angular resolution, it is possible to re-evaluate the question: do all extended discs contain substructures? We use the sample of \citet{NienkeGijs} and select all extended discs with measured 68\% continuum radii of >30 AU in the main star-forming regions: Taurus, Ophiuchus, Chamaeleon, Lupus, Upper Scorpius, Upper Centaurus-Lupus and Lower Centaurus-Crux. For Upper Sco, we have extended the sample from \citet{NienkeGijs} to the much larger sample of Upper Sco members identified by \citet{Luhman2020}, using ALMA data from e.g. \citet{Carpenter2025}. The full Upper Sco sample and its classification is described in Appendix \ref{appendix:uppersco}. For the other regions, a number of targets with 68\% disc radii >30 AU were listed in \citet{NienkeGijs} without further information on their substructures (labeled as compact or extended). This subset includes both the targets from this work, as well as from recent literature at higher angular resolution. These studies have revealed more information about these discs to reclassify them. The targets from the literature are listed in Table \ref{tab:update}, including references. Target 2MASS J04192625+2826142 was removed from the sample as it turned out to be a Class III object \citep{Rebull2010}. Target 2MASS J11432669-7804454 was marked as a 4$\sigma$ detection by \citet{Pascucci2016}, but not recovered in ALMA archival data (program 2023.1.01276.S) at higher sensitivity, so will be considered as a non-detection. We note that the full, combined survey now comprises 730 protoplanetary discs across all regions and classifications.\\
\indent Despite these updates, two discs remain classified as "Extended" without further substructure categorisation: 2MASS J11095340-7634255, which has been discussed before, and AS 205 N, which is part of a binary system and exhibits prominent spiral arms but lacks rings, gaps, or cavities (\citealt{Kurtovic2018}).\\
\begin{figure}[hbt]
    \includegraphics[scale=0.5]{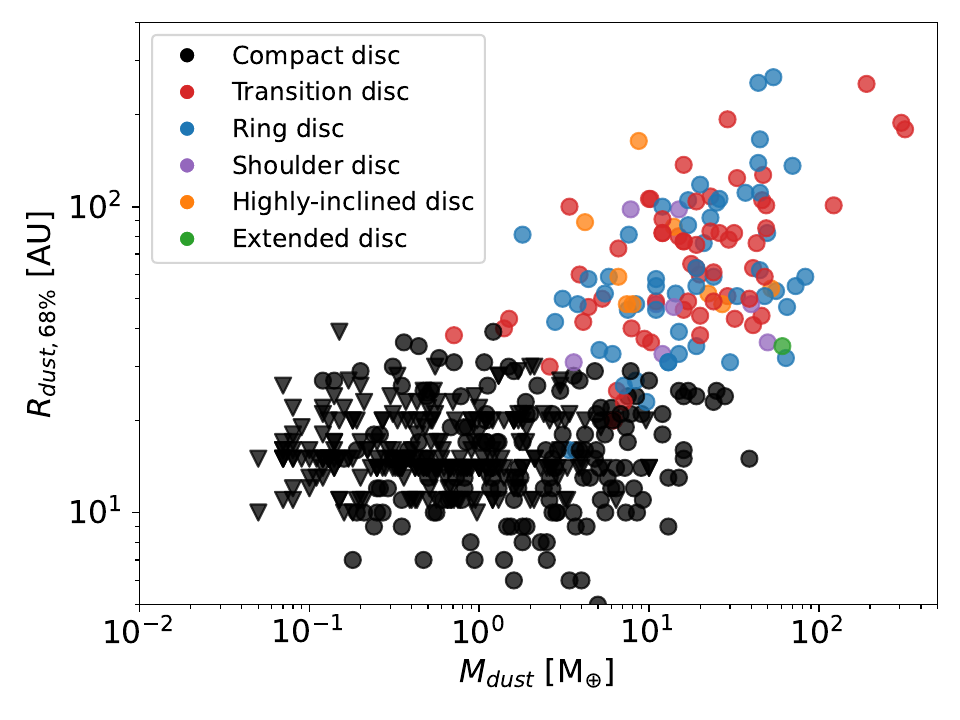}
    \caption{Dust disc size as a function of dust disc mass for all detected discs in the complete sample. Transition discs are shown in red, ring discs in blue, shoulder discs in purple, highly-inclined discs in orange, extended discs in green, and compact discs with no detected substructures in black. Circles represent detections; upper limits are indicated by triangles.}
    \label{fig:dustsize_vs_dustmass}
\end{figure}
\indent To investigate whether all extended discs contain substructures, we recreate the disc size versus dust mass diagram originally presented as Figure 1 in \citet{NienkeGijs}, now updated with our full extended disc sample and classifications (Figure \ref{fig:dustsize_vs_dustmass}). The plot confirms the hypothesis from \citet{NienkeGijs} that at least all extended discs contain large-scale substructures: for compact discs this remains mostly unknown due to lack of high-resolution data, except for the Lupus star forming region \citep{guerra2025}.\\
%The plot confirms and reinforces previous findings: transition and ring discs clearly occupy the upper-right region of the plot, consistent with them being both large and massive. In contrast, compact discs are tightly clustered in the lower-left region, indicating small dust radii and low fluxes.\\
\indent Interestingly, the shoulder and highly-inclined discs occupy the same region of the size-mass diagram as the ring and transition discs and are distinctly separated from the compact discs. This spatial segregation suggests that these discs may indeed host substructures that are either unresolved or obscured. 

\subsection{Disc morphology as a function of stellar mass, dust mass and age}
%\begin{figure}[ht]
 %   \centering
    %\includegraphics[width=0.5\textwidth]{dustmass_vs_starmass.pdf}
    %\caption{Dust mass as a function of stellar mass, grouped by region age: young (purple) and old (orange), as defined in the text. Circles denote detections, triangles indicate upper limits. Structured discs are marked as follows: transition discs ("+"), ring discs ("x"), shoulder discs ("$\Diamond$"), highly-inclined discs ("*"), and extended discs ("o"). Dashed lines show linear fits for each age group excluding structured discs, which are fit separately.}
    %\label{fig:dustmass_vs_starmass}
%\end{figure}

%\begin{table*}[ht]
%    \caption{Regression fit parameters for the M$_{dust}$-M$_{*}$ relation. Column 1 lists the groups, column 2 the intercept, column 3 the slope, column 4 the instrinsic scatter, and column 5 the associated star-forming regions.}
    %\centering 
    %\vspace{5mm}
    %\addtolength{\tabcolsep}{7pt} \renewcommand{\arraystretch}{1.4}
%    \begin{tabular}{l c c c l}
%        \hline\hline
%        \textbf{Group} & \textbf{$\alpha$ (intercept)} & \textbf{$\beta$ (slope)} & \textbf{$\delta$ (scatter)} & \textbf{Regions}\\
%        \hline
%        Young (1-5 Myr) & 0.37 $\pm$ 0.08 & 0.98 $\pm$ 0.12 & 0.39 $\pm$ 0.04 & Ophiuchus, Taurus, Lupus, Chamaeleon\\
%        Old ($\sim$10 Myr) & 0.41 $\pm$ 0.11 & 1.43 $\pm$ 0.16 & 0.30 $\pm$ 0.03 & Upper Sco, Upper Cen, Lower Cen\\
%        Structured & 1.36 $\pm$ 0.04 & 0.82 $\pm$ 0.11 & 0.14 $\pm$ 0.02 & All\\
%        \hline
%    \end{tabular}
%    \addtolength{\tabcolsep}{-7pt} 
%\label{tab:linmix}
%\end{table*}

\begin{figure}[hbt]
\centering
    %width=0.7\textwidth
    \includegraphics[scale=0.37]{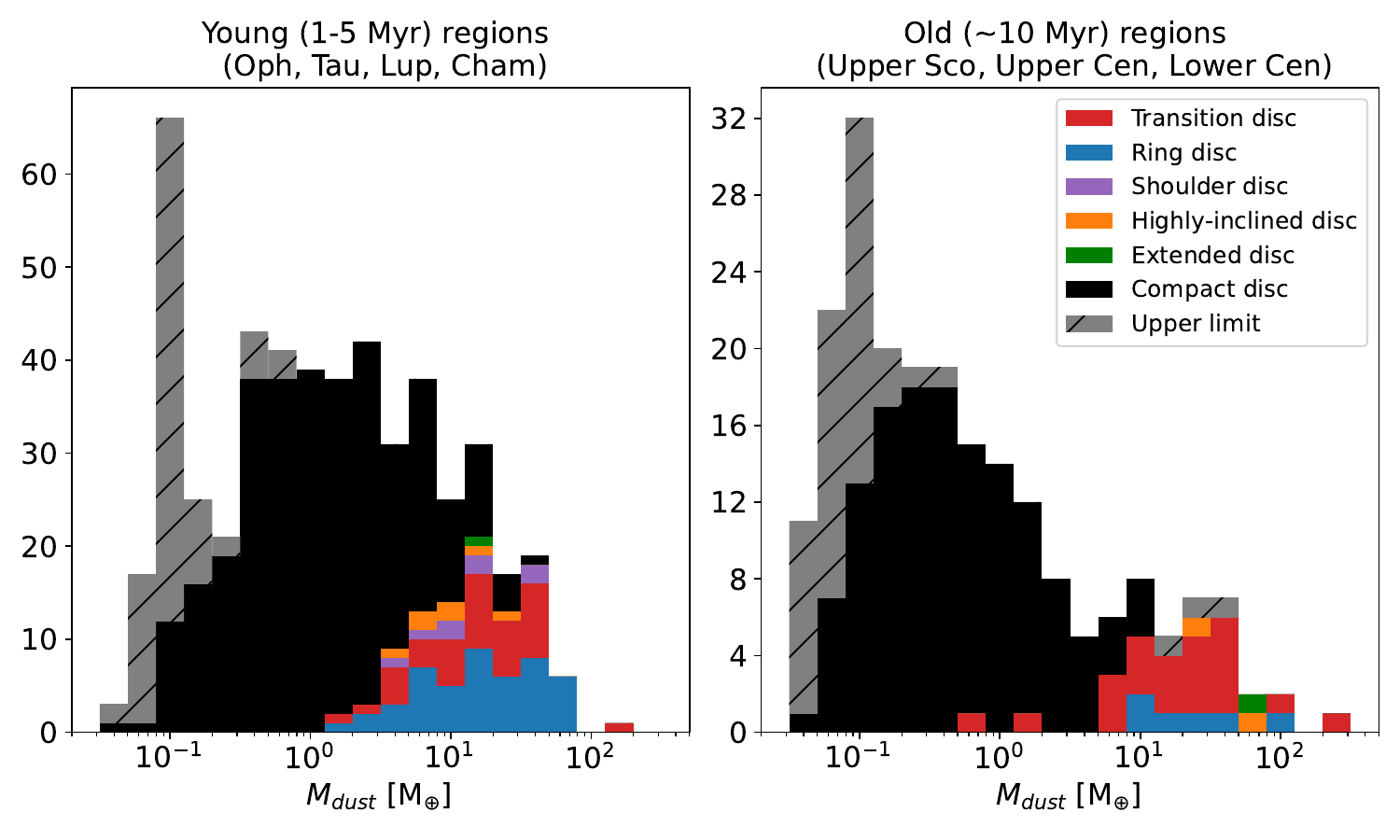}
    \caption{Distribution of dust disc masses, separated by age group. Structured discs are shown in red (transition), blue (ring), purple (shoulder), orange (highly-inclined), and green (extended). Compact discs with no detected substructures are marked in black, and upper limits are indicated in grey. For clarity, the histograms are stacked in the vertical direction and do not overlap at $y=0$.}
    \label{fig:dust_mass_distributions}
\end{figure}

\begin{figure}[hbt]
    \includegraphics[scale=0.31]{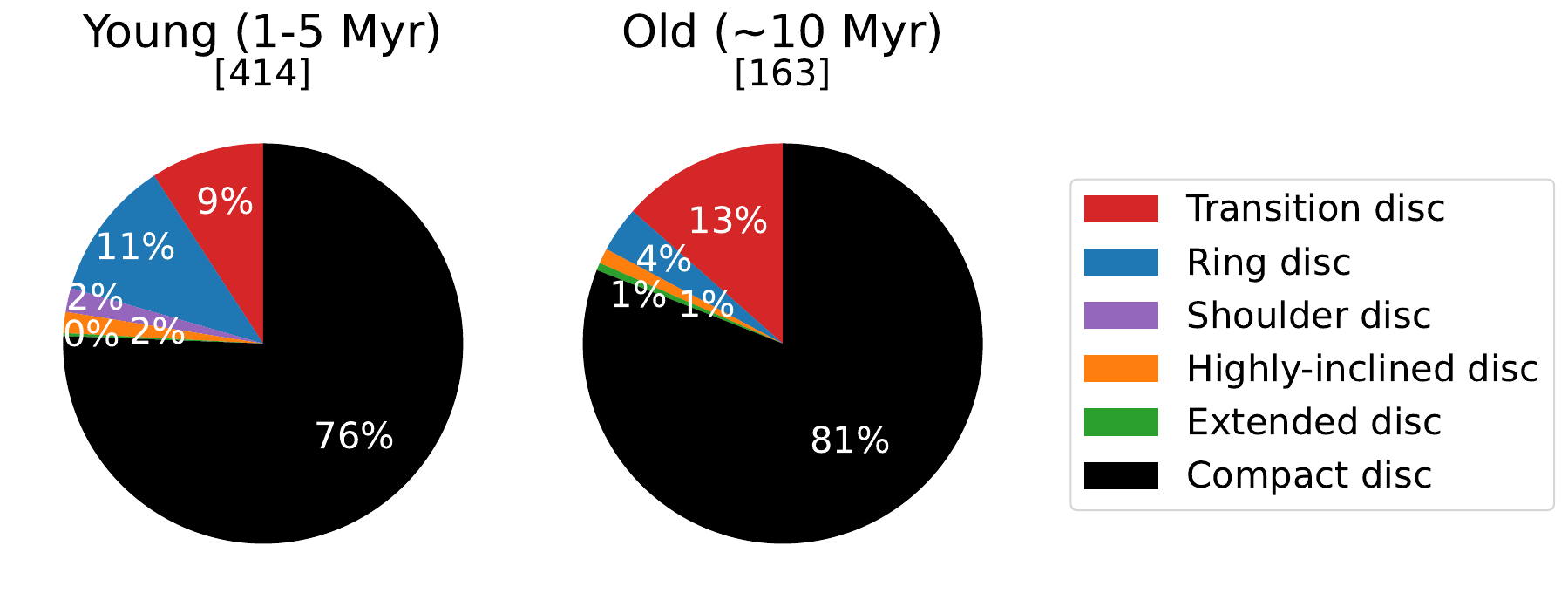}
    \caption{Distribution of disc morphologies across the two age groups. Disc types are colour-coded: red (transition), blue (ring), purple (shoulder), orange (highly-inclined), green (extended), and black (compact, no substructures). The pie charts display the percentage of each morphology within the respective age group in white. The total number of discs in each age bin is indicated in brackets above each chart.}
    \label{fig:piecharts_regions}
\end{figure}
To further investigate the evolutionary pathways of protoplanetary discs, we analyse the relationship between disc morphology, stellar mass, dust mass, and age, following the hypothesis of \citet{NienkeGijs}, using the newly found substructures as well as the updated sample of discs in Upper Sco (Appendix \ref{appendix:uppersco}). 
%Figure \ref{fig:dustmass_vs_starmass} shows the dust mass-stellar mass relation for the full disc sample, separated into two age groups: young (1-5 Myr) and old ($\sim$10 Myr). This plot updates the original Figure 4 from \citet{NienkeGijs}, now including our larger sample and the new classifications. Structured discs, defined here as transition, ring, shoulder, highly-inclined, and extended, are marked with distinct shapes, while compact discs are unmarked. Upper limits are shown as triangles. As in \citet{NienkeGijs}, we compute linear regressions for both age groups excluding structured discs, using the Bayesian technique from \citet{Kelly} implemented with the \textit{linmix} Python package. This method accounts for upper limits and intrinsic scatter. The resulting best-fit parameters (intercept $\alpha$, slope $\beta$, and intrinsic scatter $\delta$) are provided in Table \ref{tab:linmix}, following \citet{Ansdell2017}.\\
%\indent Our results confirm previous trends: disc dust mass is positively correlated with stellar mass in both age bins, and this correlation steepens with age for lower-mass stars. Structured discs are primarily located in the upper and upper-right region of the dust mass-stellar mass plane, indicating they are more common around higher-mass stars and tend to retain larger dust reservoirs over time.\\
Figure \ref{fig:dust_mass_distributions} presents the dust mass-substructure relation in histograms, separated into two age groups: young (1-5 Myr) and old ($\sim$10 Myr). Here, each class is colour-coded: transition (red), ring (blue), shoulder (purple), highly-inclined (orange), extended (green), and compact discs with no detected substructures (black), with grey representing upper limits. Structured discs are concentrated at higher dust masses ($\sim$1-200 M$_{\oplus}$), while compact discs cluster at lower values ($\sim$0.01-30 M$_{\oplus}$), with a noticeable decrease in dust mass in the older group. This supports the hypothesis of two distinct evolutionary tracks for the updated sample. Structured discs are thought to maintain higher dust masses due to the presence of pressure bumps that halt radial drift and trap millimetre-sized grains (e.g. \citealt{Pinilla2012b}; \citealt{Zhang2018}; \citealt{Dullemond2018}). Compact discs, in contrast, likely evolve without such traps and undergo rapid dust depletion, consistent with grain growth and drift models (e.g. \citealt{Birnstiel}). Based on the findings in \citet{Pinilla2025}, the decrease in dust mass seen in the compact discs may be due to leaky dust traps rather than full radial drift. It has also been suggested that compact discs may have been born small rather than a lack of strong dust traps \citep{Miotello2021}, but due to the lack of deep $^{12}$CO observations and the difficulty in their interpretation as gas disc size tracer \citep{Trapman} it has been challenging to confirm or reject this scenario.\\
\indent To further explore the occurrence of different morphologies over time, we analysed the relative fraction of disc types in young and old regions using pie charts (Figure \ref{fig:piecharts_regions}). In young regions, compact discs with no detected substructures dominate the sample (76$\pm7\%$), with ring and transition discs being the most common structured types but without significant difference in their relative occurrence (error bar computed assuming Poisson statistics). Shoulder, highly-inclined, and extended discs together make up only 3$\%$. In older regions, the fraction of compact discs with no detected substructures is 81$\pm$12$\%$, so statistically indifferent from the young regions. The difference between ring and transition disc fractions is statistically insignificant assuming Poisson errors. %, consistent with progressive dust depletion and the absence of large-scale substructures. Ring discs decline to just 4$\%$, while transition discs slightly increase to 13$\%$. The remaining types remain rare (2$\%$).\\
%\indent These trends suggest that ring structures may represent transient features that diminish or evolve over time. In contrast, the relatively stable fraction of transition discs may indicate prolonged clearing processes or continued interactions with embedded planets.\\
\begin{figure*}[hbt]
\centering
    \includegraphics[scale=0.33]{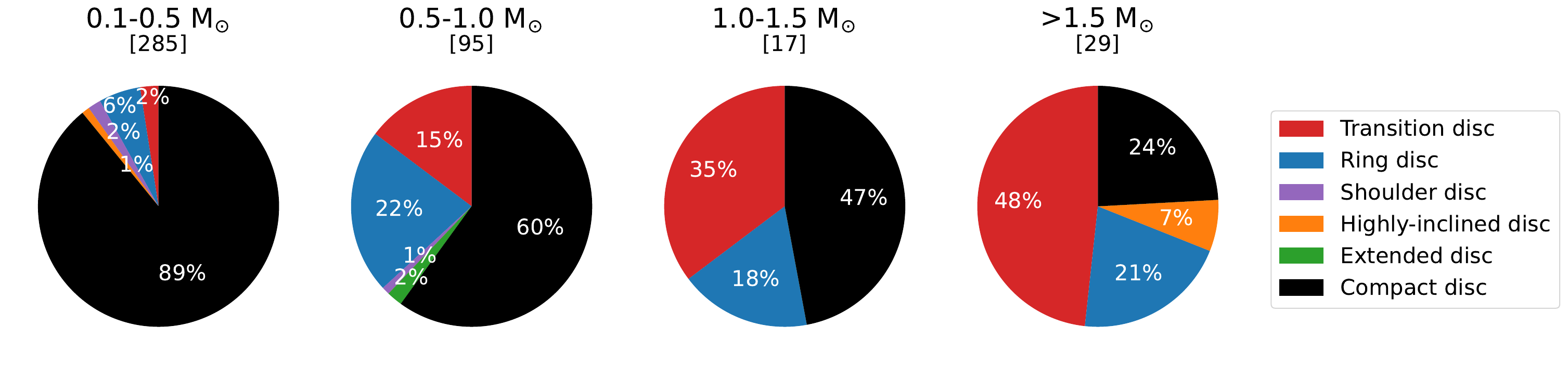}
    \caption{Distribution of disc morphologies across different stellar mass bins. Disc types are colour-coded: red (transition), blue (ring), purple (shoulder), orange (highly-inclined), green (extended), and black (compact). The pie charts display the percentage of each morphology within each stellar mass bin in white. The total number of discs in each bin is indicated in brackets above each chart.}
    \label{fig:piecharts}
\end{figure*}

\indent To assess the dependence of disc morphology on stellar mass independent of disc dust mass or age, following the same hypothesis as before, we examine the distribution of structural types across stellar mass bins (Figure \ref{fig:piecharts}). This figure updates the original Figure 6 from \citet{NienkeGijs}, presenting pie charts that show the relative fractions of transition, ring, shoulder, highly-inclined, extended, and compact discs with no detected substructures for each stellar mass category.\\
\indent The results reproduce the previously reported correlation between stellar mass and disc structure. Structured discs are significantly more prevalent around higher-mass stars. Transition discs, in particular, show a strong dependence on stellar mass: they account for only 2$\%$ of the discs in the very low-mass bin (0.1-0.5 M$_{\odot}$), increase to $\sim$18$\%$ in the intermediate-mass range (0.5-1.5 M$_{\odot}$), and peak at 48$\%$ for stars above 1.5 M$_{\odot}$. Ring discs exhibit a similar trend, rising from 6$\%$ in the lowest mass bin to 21$\%$ in the intermediate and high-mass bins.\\
\indent Interestingly, while transition discs dominate among high-mass stars, ring discs slightly outnumber them in the subsolar mass regime. Compact discs with no detected substructures remain the dominant morphology across most bins, but decrease in frequency with increasing stellar mass, from 89$\%$ in the very low-mass bin, to only 24$\%$ in the highest. These findings suggest that higher-mass stars are more likely to host structured discs, particularly transition and ring types, possibly due to more favourable conditions for substructure formation such as larger and more massive discs (e.g. \citealt{Andrews2018}; \citealt{vandermarel2018b}). Whereas this hypothesis does depend on the assumption that large-scale substructures are only detectable in larger discs \citep{Bae2023}, the results by \citet{guerra2025} do exclude large-scale substructures at wide (>30 AU) orbital radii for compact discs in Lupus, consistent with their compact dust sizes. %If those substructures are caused by planets, planet-disc interaction models do exclude giant planets at wide orbital radii in such discs, as these would have carved detectable gaps. 

%Overall, structured discs represent the majority among the highest-mass stars in the sample, comprising over two-thirds of the population in that bin.

\subsection{Compact discs}
Within the large sample, 445 discs are classified as compact (68\% radii <30 AU) and appear smooth, lacking detectable substructures at current resolutions. This is consistent with earlier studies that did not detect substructures in compact discs in protoplanetary disc surveys at typical angular resolutions (e.g. \citealt{Long2019}; \citealt{Kurtovic2021}; \citealt{vandermarel2022}; \citealt{Miley2024}; \citealt{Shi}). While most compact discs appear featureless, recent high-resolution observations have revealed substructures in a minority of them, including cavities and faint rings down to scales of $\sim$4 AU (\citealt{guerra2025}) This suggests that some substructures may remain undetected due to limited resolution or contrast, especially in the inner disc regions (\citealt{Long2019}).\\
\indent Compact discs are thought to evolve through leaky dust traps or radial drift \citep{NienkeGijs,Pinilla2025}, considering their decrease in millimetre-flux (dust mass) while retaining their typical dust disc size. %potentially depleting their small dust grains and concentrating solids toward the inner disc. 
Their location on the size-luminosity relation aligns with a drift-dominated evolution scenario (\citealt{guerra2025}), though deviations may arise due to disc truncation in binaries or the presence of unresolved dust traps. Studies analysing the gas-to-dust disc size ratio generally argue against radial drift as common feature \citep[e.g.][]{Toci2021}, but these studies do not include the fainter discs, biasing their sample. Compact discs may already have converted much of their solids into planetesimals or planets, making them promising analogues for the formation of compact planetary systems, including Super-Earths. As resolution improves, previously smooth discs increasingly show signs of substructure, suggesting that even higher-resolution observations may reveal hidden features in compact discs. 

\subsection{Do all extended discs contain substructures?}
Among the 125 extended discs in our sample ($R_{68\%}$ >30 AU), 114 exhibit resolved substructures such as rings, gaps, cavities, or shoulders. The remaining 11 include 9 highly-inclined systems, where detection is limited by projection effects, and 2 ambiguous cases: 2MASS J11095340-7634255 (likely a superimposed binary), and AS 205 N (part of a binary system with spiral-like emission but no clear rings or gaps). This yields a raw detection rate of 91\% (114/125), representing the fraction of extended discs in the sample where substructures are observed. However, excluding the 9 highly-inclined discs, where substructures may be present but are not detectable due to projection effects, the detection rate among discs where substructures can be reliably identified rises to $\sim$98$\%$ (114/116).\\
\indent Extended, apparently smooth discs are also rare in other regions. MP Mus ($R_{68\%}$ = 30 AU), originally classified as smooth by \citet{Ribas}, was recently found to host additional gaps and an inner cavity based on new high-resolution observations at longer wavelengths (\citealt{Aguayo,Ribas2025}).\\
\indent Taken together, these results suggest that substructures are a near-universal feature of extended discs. The few exceptions may be attributed to observational limitations, such as inclination, or to physical effects like binarity and dynamical truncation. The ubiquity of substructures has important implications for dust evolution: pressure maxima associated with rings and gaps can trap particles, slow radial drift, and promote grain growth (\citealt{Pinilla2012b}). Such dust traps may also serve as favourable sites for planetestimal and ultimately planet formation, potentially resolving the radial drift and fragmentation barriers in planet formation theory (e.g. \citealt{Birnstiel2012}; \citealt{Carrera2015}). The early emergence and widespread occurrence of substructures therefore appear to play a central role in shaping the initial conditions for planet formation.

\section{Conclusion}
\label{sec:Conclusion}
We present high-resolution ALMA Band 6 (1.33 mm)  observations at $\sim$0.12" resolution of 26 protoplanetary discs, primarily focusing on extended discs with 68\% dust radii larger than 30 AU. These observations complete the known sample of nearby discs in the main star-forming regions and aim to address the question: Do all extended discs contain substructures? We compare the full disc sample and its classifications with stellar age and mass. The key findings of this work are:

\begin{itemize}
%    \item We obtained $\sim$0.12" resolution images of 26 discs and analysed their dust continuum emission using \textit{Frankenstein} and \textit{Galario}.
    \item Of the 26 analysed discs, 17 show substructures: six are transition discs, 6 are ring discs, and 5 are shoulder discs.
    \item Nine discs lack clear substructures: six are compact ($R_{68\%}$ <30 AU), two are highly inclined, and one (2MASS J11095340-7634255) shows an irregular morphology possibly due to a superimposed binary or late-stage infall.
    \item Discs with substructures have 68\% effective dust radii between 23 and 98 AU.
    \item We detect $^{12}$CO J=2-1 emission in 15 discs. In all cases, the gas discs are larger than the dust discs, consistent with the higher opacity of $^{12}$CO rotational emission compared to millimetre continuum emission. Whether this is also a sign of radial drift requires more detailed analysis \citep{Trapman2020}.
    %\item Keplerian rotation is observed in most discs, with moment 1 maps showing clear velocity gradients.
    \item Extended CO structures are observed in four discs, including the known tail in SU Aur, suggesting ongoing accretion or late infall; additional tails in other discs may relate to outflows or remnant envelope material.
%    \item Analysis of the complete sample reveals a positive correlation between dust mass and stellar mass.
    \item Building on the hypothesis by \citet{NienkeGijs}, we have analysed a large sample of discs by dividing it in structured discs and smooth, compact discs with $R_{68\%}$ <30 AU. We reproduce the previous findings that the latter category decreases in dust mass, whereas structured discs do not, with an updated sample (Figure \ref{fig:dust_mass_distributions}). With the same classification, we reproduce a previous result on the stellar mass dependence of structured discs, where structured discs appear to be more common around higher-mass stars. Transition discs show a strong stellar mass dependence, increasing in occurrence from 2\% in the 0.1–0.5 M$_{\odot}$ range to 48\% for stars above 1.5 M$_{\odot}$.
    \item Nearly all ($\sim$91\%) extended discs ($R_{68\%}$ >30 AU) contain substructures; when accounting for highly-inclined systems where substructures may be undetectable, the fraction may be as high as $\sim$98\%.
    \item The prevalence of substructures in extended discs and their correlation with stellar mass imply that the origin of substructures could have a stellar mass dependence as well. This is consistent with the proposed hypothesis by \citet{NienkeGijs}, that substructures are dust traps caused by giant planets at wide orbital radii, as exoplanets show a similar stellar mass dependence. 
    %that dust evolution is efficient and possibly accelerated in massive systems. 
    %Substructures may act as dust traps, slowing radial drift and promoting grain growth. These processes are critical for planetesimal formation, suggesting that the presence and nature of substructures play a key role in shaping planet formation pathways.
\end{itemize}

\begin{acknowledgements}
This paper makes use of the following ALMA data: ADS/JAO.ALMA\#2022.1.01302.S. ALMA is a partnership of ESO (representing its member states), NSF (USA) and NINS (Japan), together with NRC (Canada), NSTC and ASIAA (Taiwan), and KASI (Republic of Korea), in cooperation with the Republic of Chile. The Joint ALMA Observatory is operated by ESO, AUI/NRAO and NAOJ. G.D.M. acknowledges support from FONDECYT project 1252141 and the ANID BASAL project FB210003. The PI acknowledges assistance from Allegro, the European ALMA Regional Center node in the Netherlands. We thank Melissa McClure for useful discussions.
\end{acknowledgements}

\bibliographystyle{aa.bst}
\bibliography{ReferencesPaper.bib}

\begin{appendix}
\clearpage
\section{Sample and observational details}
\label{appendix:host_star_disc_properties_table}
This appendix provides supplementary information on the target sample and the ALMA observations used in this study. Table \ref{tab:properties} summarises the stellar properties, dust disc radii, and estimated dust disc masses of the 26 protoplanetary discs, taken from literature. Table \ref{tab:observations} lists the key observational parameters for the 23 star systems, including beam sizes, integration times, weather conditions, and calibrators.
\begin{table*}[hb]
    \caption{The host star properties and estimated dust disc sizes and dust disc masses of the 26 protoplanetary discs taken from literature. Column 1 lists the source names, and column 2 indicates the associated star-forming regions (with "Upper Cen" and "Upper Sco" representing Upper Centaurus-Lupus and Upper Scorpius, respectively). Columns 3–5 provide the distances (pc), spectral types, and stellar masses (M$_\odot$). Column 6 shows the estimated 68\% dust disc radii (AU), with upper limits denoted by "<". Column 7 lists the dust disc masses (M$_{\text{Earth}}$), and column 8 includes the references for all listed values. Full citations are provided below the table.
}
    \centering 
    %\vspace{5mm}
    %\renewcommand{\arraystretch}{1.5}
    \begin{tabular}{l c c c c c c c c}
        \hline\hline
        \textbf{Disc Name} & \textbf{Region} & \textbf{Dist} & \textbf{SpT} & \textbf{\textit{M$_{*}$}} & \textbf{\textit{R$_{\textbf{dust, 68\%}}$}} & \textbf{\textit{M$_{\textbf{dust}}$}} & \textbf{Reference}\\
        & & \textbf{[pc]} & & \textbf{[M$_{\odot}$]} & \textbf{[AU]} & \textbf{[M$_{\oplus}$]}& \\
        \hline
        2MASS J11095340-7634255 & Chamaeleon & 193 & K7 & 0.7 & 56 & 13 & a, b, c\\
        2MASS J11004022-7619280 & Chamaeleon & 193 & M4 & 0.2 & 36 & 11 & a, b, c\\
        2MASS J11111083-7641574 & Chamaeleon & 185 & M1 & - & 68 & 7.8 & a, b\\
        2MASS J11160287-7624533 & Chamaeleon & 185 & K8 & - & 39 & 1.8 & a, b\\
        2MASS J11104959-7717517 & Chamaeleon & 187 & M2 & 0.4 & 31 & 8.4 & a, b, c\\
        2MASS J11094742-7726290 & Chamaeleon & 191 & M1 & 0.5 & 72 & 23 & a, b, c\\
        SZ 133 & Lupus & 116 & K5 & - & 104 & 6.6 & b, c, d \\
        RX J1556.1-3655 & Lupus & 158 & M1 & 0.5 & 42 & 6.1 & b, c, d, e\\
        SSTc2d J162652.0-243039 & Ophiuchus & 135 & M5 & - & 51 & 2.0 & f\\
        SSTc2d J162546.6-242336 & Ophiuchus & 135 & - & - & 48 & 4.2 & f\\
        SSTc2d J162718.4-243915 & Ophiuchus & 135 & - & - & 35 & 5.5 & f\\
        SSTc2d J162738.3-235732 & Ophiuchus & 141 & K6 & - & <14 & - & c, f\\
        SSTc2d J162145.1-234232 & Ophiuchus & 135 & - & - & 36 & 7.2 & f\\
        SSTc2d J162823.3-242241 & Ophiuchus & 147 & K5 & - & 62 & 3.9 & c, f\\
        SSTc2d J163952.9-241931 & Ophiuchus & 135 & -  & - & 31 & 1.5 & f\\
        Haro 6-37 A & Taurus & 206 & K8 & 0.6 & 13 & 0.8 & c, g\\
        Haro 6-37 B & Taurus & 195 & M1 & 0.4 & 34 & 15 & c, g\\
        MHO 1 & Taurus & 134 & M2.5 & 0.3 & 26 & 40 & c, g, h\\
        MHO 2 & Taurus & 131 & M2.5 & 0.7 & 37 & 24 & c, g, h\\
        IT Tau A & Taurus & 160 & K6 & 0.7 & <32 & 1.9 & c, g, h\\
        IT Tau B & Taurus & 171 & M2.9 & 0.3 & <32 & 1.1 & c, g, h\\
        SU Aur & Taurus & 157 & G4 & 3.1 & 50 & 7.1 & c, g\\
        CY Tau & Taurus & 126 & M2.3 & 0.3 & 48 & 14 & c, g, i\\
        2MASS J04154278+2909597 & Taurus & 160 & M1.3 & 0.4 & 39 & 5.3 & c, g, i\\
        HD 139614 & Upper Cen & 134 & A9 & 1.6 & 67 & 46 & c, j, k\\
        2MASS J16075796-2040087 & Upper Sco & 137 & M1 & 0.5 & <34 & 3.9 & c, h\\
        
        \hline
    \end{tabular}
    \caption*{\textbf{References.} (a) \citet{Pascutti}, (b) \citet{Manara}, (c) \citet{Gaia}, (d) \citet{Alcal}, (e) \citet{Ansdell}, (f) \citet{Cieza}, (g) \citet{Akeson}, (h) \citet{Barenfeld}, (i) \citet{tripathi}, (j) \citet{Ansdell2020}, (k) \citet{Vioque}.}
\label{tab:properties}
\end{table*}

\begin{table*}[ht]
    \caption{Observational details of the 23 targets. Column 1 lists the target names, and column 2 provides the beam size (arcseconds). Columns 3 and 4 show the observation dates and integration times (seconds). Column 5 indicates the maximum recoverable scales (arcseconds) based on the baselines. Column 6 gives the average precipitable water vapour (PWV) in millimetres. Column 7 lists the calibrators, where the first calibrator is used for both bandpass and flux, and the second is for the phase calibration.}
    %\hspace{-16mm}
    \centering 
    \begin{tabular}{l c c c c c c}
        \hline\hline
        \textbf{Target Name} & \textbf{Beam Size} & \textbf{Date} & \textbf{Int. Time} & \textbf{Max. Reco. Scale} &\textbf{PWV} & \textbf{Calibrators} \\
        & \textbf{["]} & & \textbf{[min]} & \textbf{["]} & \textbf{[mm]} &\\
        \hline
        2MASS J11095340-7634255 & $0.180\times0.104$ & 25-05-2023 & 12 & 1.87 & 0.40 & J1427-4206, J1058-8003\\
        & $0.745\times0.406$ & 16-01-2023 & 3 & 7.59 & 0.96 & J0519-4546, J058-8003\\
        2MASS J11004022-7619280 & $0.180\times0.104$ & 25-05-2023 & 12 & 1.87 & 0.40 & J1427-4206, J1058-8003\\
        & $0.745\times0.406$ & 16-01-2023 & 3 & 7.53 & 0.96 & J0519-4546, J058-8003\\
        2MASS J11111083-7641574 & $0.180\times0.104$ & 25-05-2023 & 12 & 1.88 & 0.40 & J1427-4206, J1058-8003\\
        & $0.745\times0.406$ & 16-01-2023 & 3 & 7.59 & 0.96 & J0519-4546, J058-8003\\
        2MASS J11160287-7624533 & $0.180\times0.104$ & 25-05-2023 & 12 & 1.86 & 0.40 & J1427-4206, J1058-8003\\
        & $0.745\times0.406$ & 16-01-2023 & 3 & 7.61 & 0.96 & J0519-4546, J058-8003\\
        2MASS J11104959-7717517 & $0.167\times0.107$ & 29-05-2023 & 12 & 2.03 & 0.76 & J1427-4206, J1058-8003\\
        & $0.829\times0.428$ & 16-01-2023 & 3 & 7.75 & 1.02 & J0519-4546, J058-8003\\
        2MASS J11094742-7726290 & $0.167\times0.107$ & 29-05-2023 & 12 & 2.04 & 0.76 & J1427-4206, J1058-8003\\
        & $0.829\times0.428$ & 16-01-2023 & 3 & 7.78 & 1.02 & J0519-4546, J058-8003\\
        SZ 133 & $0.147\times0.112$ & 17-05-2023 & 11 & 2.28 & 1.35 & J1427-4206, J1610-3958\\
        & $0.936\times0.647$ & 02-01-2023 & 3 & 7.74 & 1.70 & J1427-4206, J1610-3958\\
        RX J1556.1-3655 & $0.122\times0.105$ & 24-05-2023 & 10 & 1.44 & 0.51 & J1427-4206, J1610-3958\\
        SSTc2d J162652.0-243039 & $0.152\times0.124$ & 17-05-2023 & 10 & 2.16 & 1.34 & J1427-4206, J1617-2537\\
        SSTc2d J162546.6-242336 & $0.152\times0.124$ & 17-05-2023 & 10 & 2.14 & 1.34 & J1427-4206, J1617-2537\\
        SSTc2d J162718.4-243915 & $0.152\times0.124$ & 17-05-2023 & 10 & 2.14 & 1.34 & J1427-4206, J1617-2537\\
        SSTc2d J162738.3-235732 & $0.152\times0.124$ & 17-05-2023 & 10 & 2.17 & 1.34 & J1427-4206, J1617-2537\\
        SSTc2d J162145.1-234232 & $0.161\times0.117$ & 17-05-2023 & 10 & 2.33 & 1.19 & J1924-2914, J1617-2537\\
        SSTc2d J162823.3-242241 & $0.152\times0.124$ & 17-05-2023 & 10 & 2.14 & 1.34 & J1427-4206, J1617-2537\\
        SSTc2d J163952.9-241931 & $0.140\times0.115$ & 16-05-2023 & 10 & 1.82 & 1.69 & J1427-4206, J1700-2610\\
        & $0.821\times0.632$ & 02-01-2023, & 5 & 7.43 & 1.22 & J1427-4206, J1625-2527\\
        & & 03-01-2023 & & &\\
        Haro 6-37 A+B & $0.114\times0.107$ & 02-06-2023 & 12 & 1.84 & 0.57 & J0423-0120, J0440+1437\\
        MHO 1+2 & $0.192\times0.104$ & 31-05-2023 & 15 & 2.01 & 0.96 & J0423-0120, J0438+3004\\
        IT Tau A+B & $0.139\times0.103$ & 30-05-2023 & 14 & 1.85 & 1.22 & J0423-0120, J0438+3004\\
        SU Aur & $0.161\times0.099$ & 23-05-2023 & 12 & 1.89 & 0.43 & J0423-0120, J0438+3004\\
        CY Tau & $0.192\times0.104$ & 31-05-2023 & 15 & 2.01 & 0.96 & J0423-0120, J0438+3004\\
        2MASS J04154278+2909597 & $0.152\times0.102$ & 21-05-2023, & 30 & 2.12 & 0.48 & J0423-0120, J0438+2600\\
        & & 23-05-2023 & & &\\
        HD 139614 & $0.157\times0.112$ & 17-05-2023 & 11 & 2.16 & 1.37 & J1427-4206, J1604-4441\\
        & $0.675\times0.568$ & 07-01-2023 & 3 & 5.88 & 1.27 & J1427-4206, J1604-4441\\
        2MASS J16075796-2040087 & $0.121\times0.086$ & 21-05-2023, & 12 & 1.45 & 0.90 & J1427-4206, J1551-1755\\
        & & 22-05-2023 & & &\\
        
        \hline
    \end{tabular}
\label{tab:observations}
\end{table*}

\section{$^{12}$CO gas results}
\label{appendix:12 CO moment maps}
\begin{figure*}[hb]
    %\hspace{-12mm}
    \centering
    \includegraphics[scale=0.2]{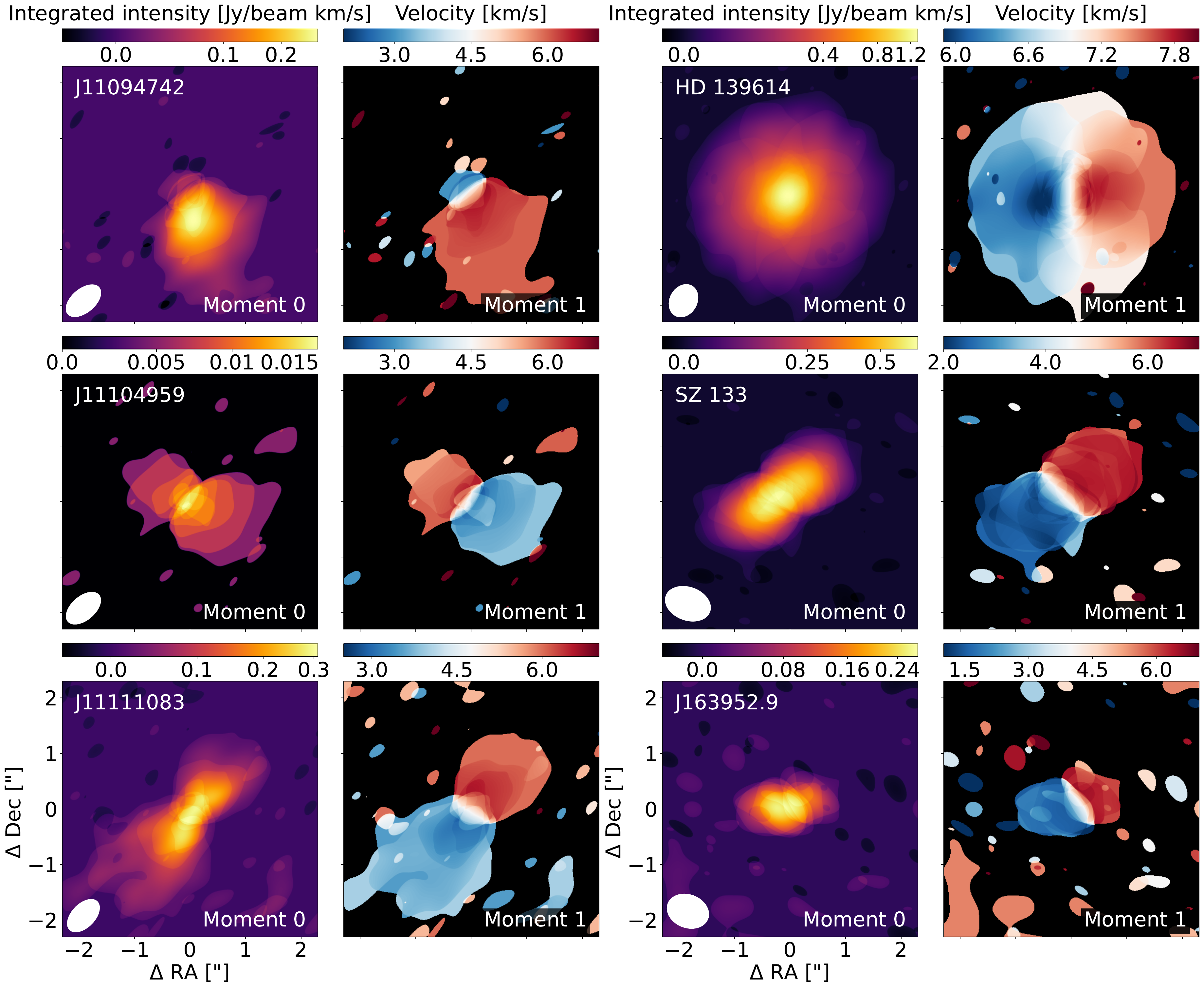}
    \caption{Gallery of the low-resolution $^{12}$CO J=2-1 moment maps for 2MASS J11094742-7726290, 2MASS J11104959-7717517, 2MASS J11111083-7641574 (left panels, top to bottom), HD 139614, SZ 133, and SSTc2d J163952.9-241931 (right panels, top to bottom), constructed from emission exceeding the 3$\sigma$ threshold. Moment 0 maps (left columns) display integrated intensity in Jy/beam km/s, and moment 1 maps (right columns) show velocity in km/s. Axes indicate RA and Dec offsets from the disc centre in arcseconds. Each panel spans $4.6\text{"}\times4.6\text{"}$, with white ellipses in moment 0 maps representing beam sizes. For discs with names starting with SSTc2d or 2MASS, the names in the upper left corners of the moment 0 maps are abbreviated to the first part after "J" (e.g., 2MASS J11094742-7726290 is labelled as J11094742).}
    \label{fig:CO_low}
\end{figure*}
\begin{figure*}[ht]
    %\hspace{-12mm}
    \centering
    \includegraphics[scale=0.2]{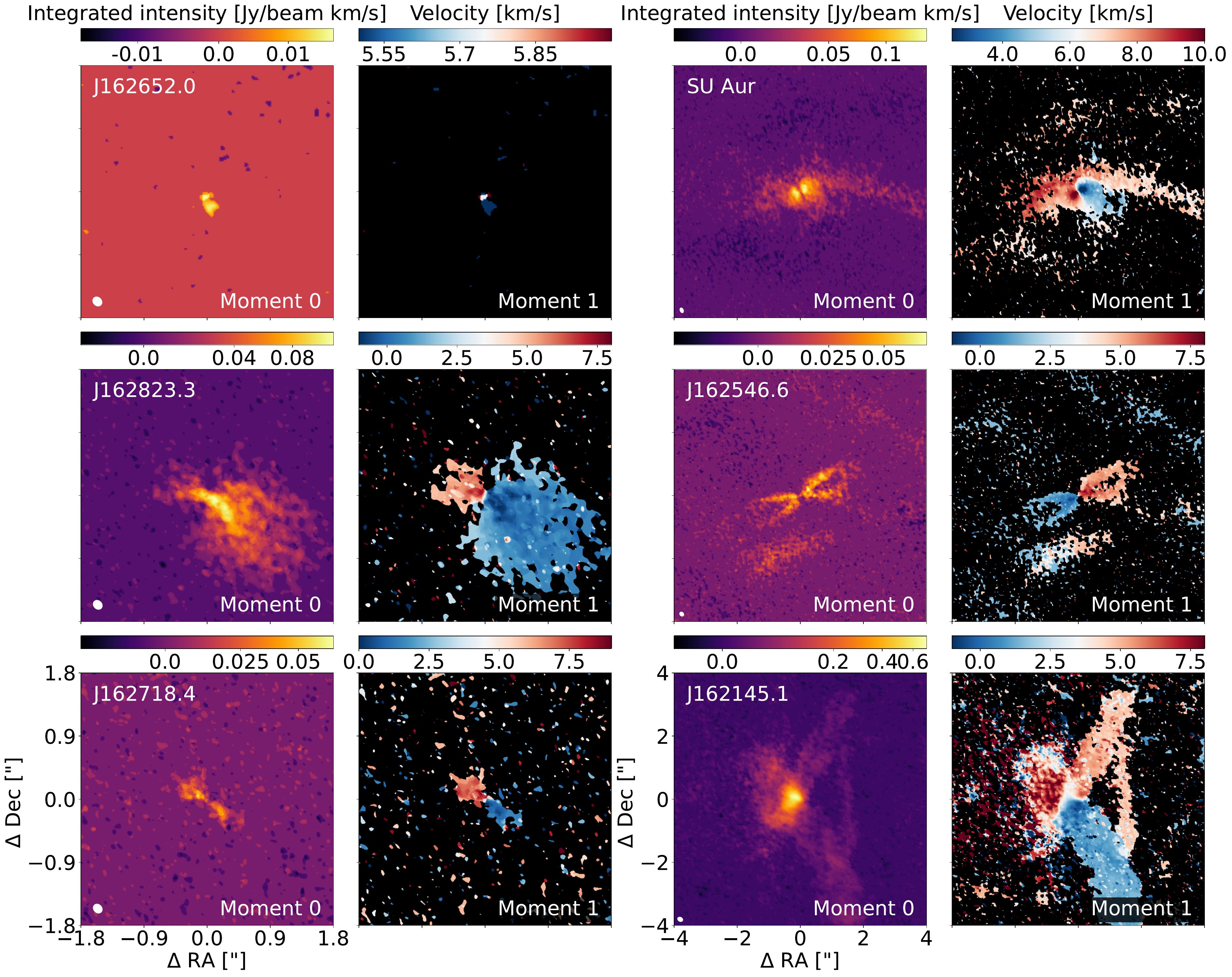}
    \caption{Gallery of the low-resolution $^{12}$CO J=2-1 moment maps for SSTc2d J162652.0-243039, SSTc2d J162823.3-242241, SSTc2d J162718.4-243915 (left panels, top to bottom), SU Aur, SSTc2d J162546.6-242336, and SSTc2d J162145.1-234232 (right panels, top to bottom), constructed from emission exceeding the 3$\sigma$ threshold. Moment 0 maps (left columns) display integrated intensity in Jy/beam km/s, and moment 1 maps (right columns) show velocity in km/s. Axes indicate RA and Dec offsets from the disc centre in arcseconds. The left panels span $3.6\text{"}\times3.6\text{"}$, while the right panels span $8.0\text{"}\times8.0\text{"}$. White ellipses in moment 0 maps representing beam sizes. For discs with names starting with SSTc2d or 2MASS, the names in the upper left corners of the moment 0 maps are abbreviated to the first part after "J" (e.g., SSTc2d J162652.0-243039 is labelled as J162652.0).}
    \label{fig:CO_high}
\end{figure*}
\begin{figure}[ht]
    \centering
    \includegraphics[scale=0.27]{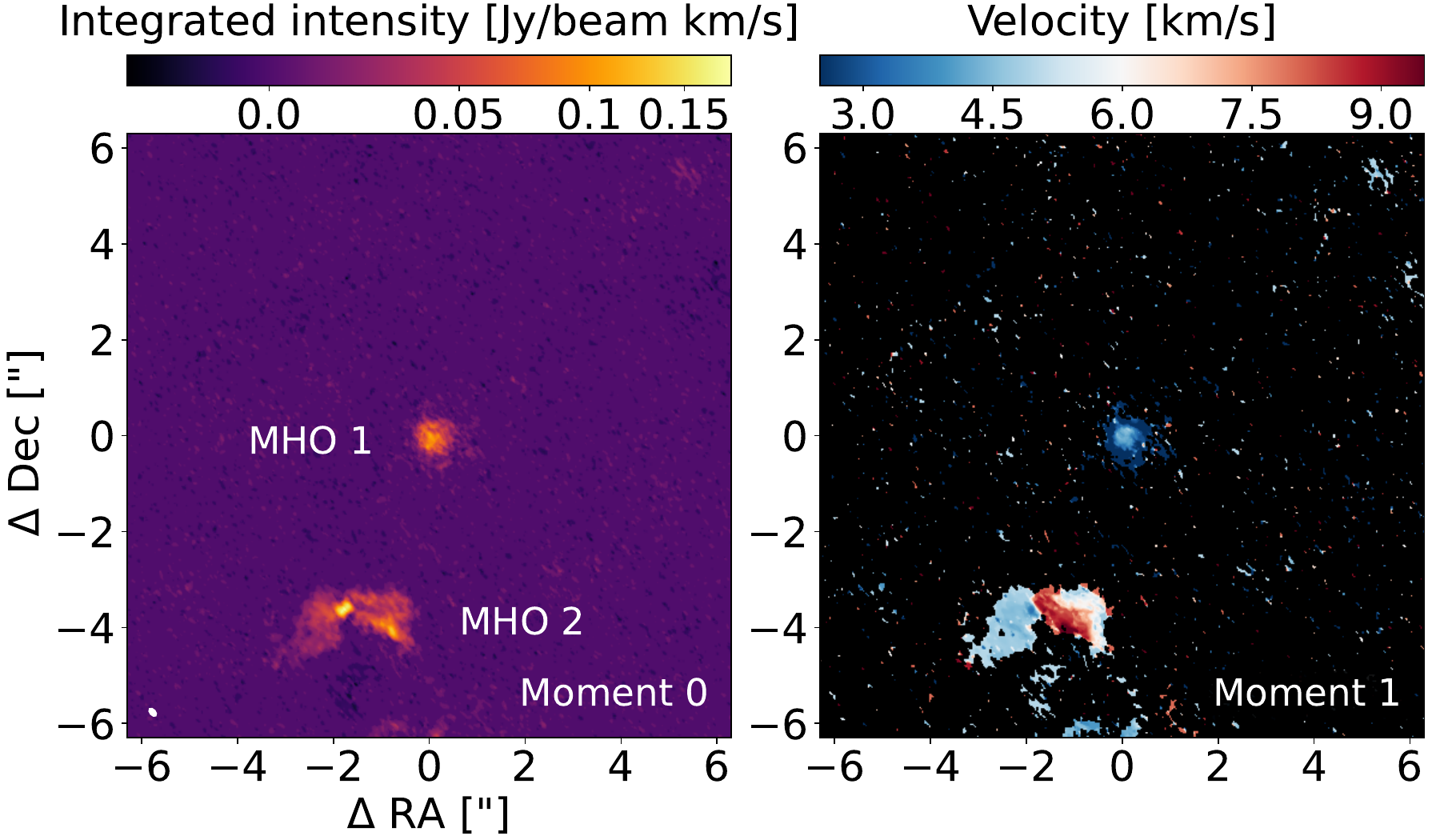}
    \caption{High-resolution $^{12}$CO J=2-1 moment maps for MHO 1+2, constructed from emission exceeding the 3$\sigma$ threshold. The moment 0 map (left) displays integrated intensity in Jy/beam km/s, and the moment 1 map (right) shows velocity in km/s. Axes indicate RA and Dec offsets from the disc centre in arcseconds. Both panels span $12.6\text{"}\times12.6\text{"}$, with a white ellipse in the moment 0 map representing the beam size.}
    \label{fig:CO_binary}
\end{figure}
We detected $^{12}$CO J=2-1 line emission in 15 discs: 2MASS J11095340-7634255, 2MASS J11094742-7726290, 2MASS J11104959-7717517, 2MASS J11111083-7641574, HD 139614, SZ 133, SSTc2d J163952.9-241931, SSTc2d J162652.0-243039, SSTc2d J162823.3-242241, SSTc2d J162718.4-243915, SU Aur, SSTc2d J162546.6-242336, SSTc2d J162145.1-234232, and the binary system MHO 1 and MHO 2. For the first seven discs, low-resolution data was available and used to create moment maps, while for the remaining eight discs, only high-resolution data was used. We first discuss the results from the low-resolution maps, followed by the high-resolution maps. The binary system is addressed separately, as the gas inside of these discs may dynamically influence one another. Additionally, we discuss 2MASS J11095340-7634255 individually due to its unique dust emission morphology.
\subsection{Low-resolution $^{12}$CO moment maps}
Figure \ref{fig:CO_low} presents the $^{12}$CO moment 0 and moment 1 maps for the six discs observed at low angular resolution. The moment 0 maps reveal no discernible substructures, with the CO gas emission generally aligning with the dust continuum emission. However, the gas discs consistently extend beyond their corresponding dust discs. This size disparity is attributed to the higher optical depth of CO gas compared to dust, allowing as to remain detectable at greater distances (e.g. \citealt{Dutrey}; \citealt{Guilloteau}; \citealt{Facchini}), as well as radial drift concentrating dust grains in the inner disc regions (e.g. \citealt{Testi}; \citealt{Natta}; \citealt{Birnstiel}).\\
\indent Foreground absorption, likely originating from nearby cold molecular clouds, is observed in the channel maps of nearly all discs. This absorption, particularly pronounced in 2MASS J11094742-7726290, introduces asymmetries in the moment maps. Such foreground absorption is common in star-forming regions like Ophiuchus, where molecular clouds with low temperatures and narrow intrinsic line widths shield CO disc emission in specific velocity ranges (e.g. \citealt{Boogert}).\\
\indent The moment 1 maps show distinct redshifted (red) and blueshifted (blue) regions on opposite sides of each discs, characteristic of Keplerian rotation. Similarly to other ALMA disc observations (e.g. \citealt{Ansdell}), the gas discs are larger than their dust counterparts.

\subsection{High-resolution $^{12}$CO moment maps}
Figure \ref{fig:CO_high} displays the $^{12}$CO moment 0 and moment 1 maps for the six single discs observed at high angular resolution. These maps show increased noise, evident as background speckling, compared to the low-resolution maps in Figure \ref{fig:CO_low}. This heightened noise results from the finer spatial details captured and instrumental factors, such as thermal and electronic noise associated with high-resolution observations.\\
\indent As in the low-resolution maps, no discernible substructures are visible in the moment 0 maps, and the gas disc orientation generally align with the dust continuum. Foreground absorption again introduces asymmetries, particularly in SSTc2d J162823.3-242241. The moment 1 maps show clear Keplerian rotation in five discs, indicating smooth elliptical orbits of gas particles governed by the central star's gravity. In contrast, $^{12}$CO gas in SSTc2d J162652.0-243039 is only faintly visible in a few channel maps, suggesting that foreground molecular clouds heavily obscured its emission, resulting in a weak and non-Keplerian signal.\\
\indent Three discs, SU Aur, SSTc2d J162546.6-242336, and
SSTc2d J162145.1-234232, exhibit extended CO emission beyond their protoplanetary discs. For SU Aur, the extended emission appears as a tail originating from the disc centre and extending to the right. Similar structures have been observed in optical, near-infrared scattered light, and CO studies (\citealt{Grady}; \citealt{Chakraborty}; \citealt{Bertout}; \citealt{Jeffers}; \citealt{deleon}; \citealt{Akiyama_su_aur}; \citealt{Ginski}), suggesting active accretion and a phase of late infall, potentially impacting the disc's evolution.\\
\indent SSTc2d J162546.6-242336 displays excess CO emission both below and to the upper right of the disc, possibly representing remnant material from star formation or outflows from the central star. Further research is needed to clarify its origin.\\
\indent In SSTc2d J162145.1-234232, two large CO gas tails are visible, likely influenced by protostellar outflows. \citet{Sacco} detected Ne II emission in this disc, attributing it to shock ionisation from outflows. A similar process could enhance CO emission, as shocks compress and heat the surrounding molecular gas. This may explain the extended structures seen in our observations, though further studies are needed to confirm this interpretation.

\subsection{High-resolution $^{12}$CO moment maps for MHO 1+2}
Figure \ref{fig:CO_binary} shows the $^{12}$CO moment 0 and moment 1 maps for the binary system MHO 1+2 observed at high angular resolution. In the upper disc, MHO 1, the CO emission is uniformly distributed, aligning with the dust disc, but lacks a velocity gradient or evidence of Keplerian rotation. In contrast, the lower disc, MHO 2, exhibits a clear velocity gradient in the moment 1 map, indicative of Keplerian rotation. Notably, the gas disc in MHO 2 shows no cavity corresponding to the structure seen in its dust disc, indicating that the gas cavity may be too small to be resolved.\\
\indent The absence of Keplerian rotation in MHO 1 may result from gravitational interactions with MHO 2, where tidal forces and perturbations disrupt the expected motion (see \citealt{Tofflemire}). Additionally, no extended gas emission is observed in the moment maps, suggesting the absence of a circumbinary disc in this system.   

\section{Best-fit models}
\label{appendix:Best-fit models}
\begin{figure*}[ht]
    %\hspace{-12mm}
    \centering
    \includegraphics[scale=0.175]{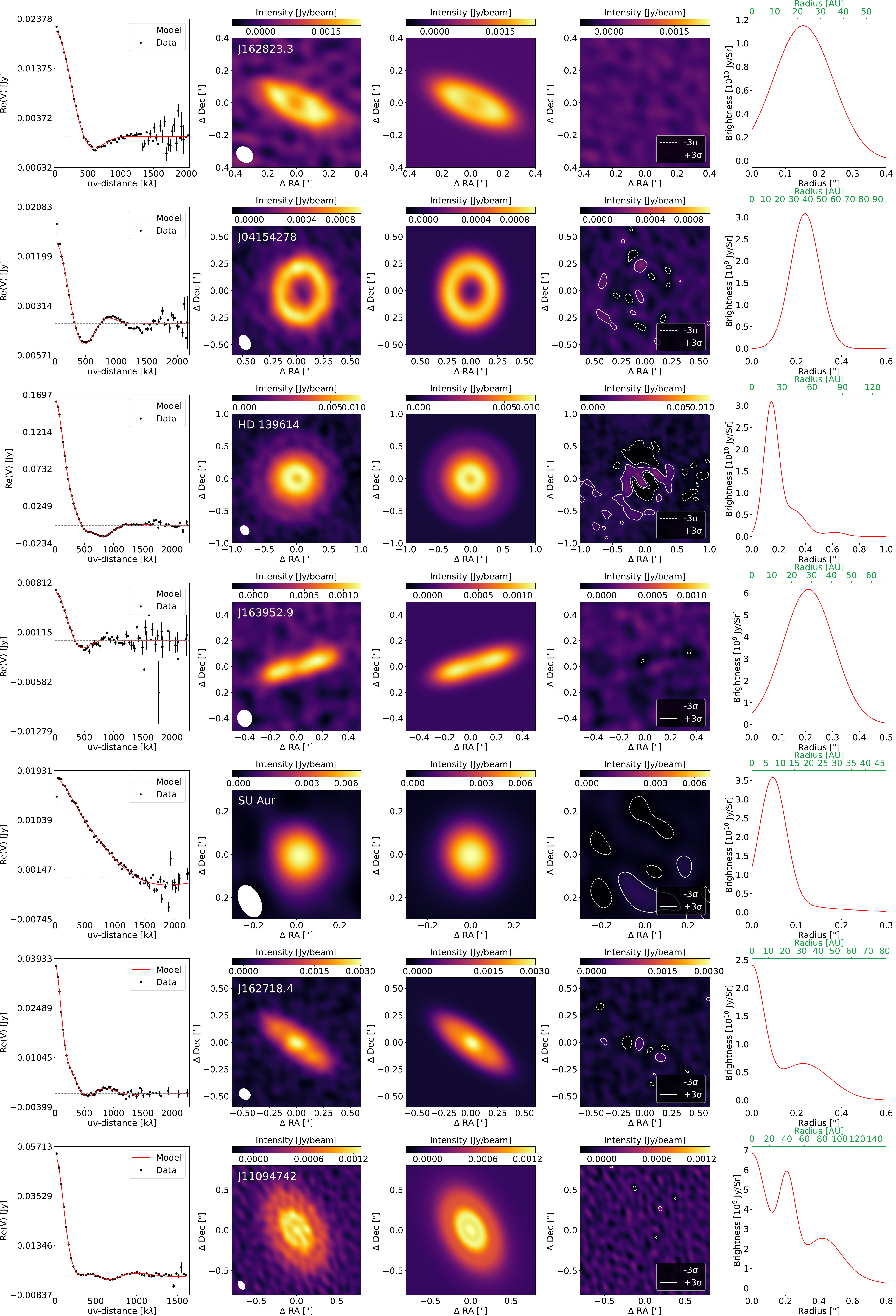}
    \caption{Comparison of data and best-fit model for the first seven modelled single discs. Column 1 shows deprojected and binned visibilities (observations in black, models in red) as a function of uv-distance. Columns 2 and 3 display the cleaned data and model images, convolved with the same beam. For discs with names starting with SSTc2d, 2MASS, or RX, the names in the upper left corners of column 2 are abbreviated to the first part after "J" (e.g., SSTc2d J162823.3-242241 is labelled as J162823.3). Column 4 presents residual images with $\pm$3$\sigma$ contours in white. RA and Dec offsets are in arcseconds, and the colour scale represents intensity in Jy/beam. Beam sizes are indicated in the lower left. Column 5 shows the radial brightness profiles with AU conversions in green above each profile. }
    \label{fig:galario_first}
\end{figure*}

\begin{figure*}[ht]
    %\hspace{-12mm}
    \centering
    \includegraphics[scale=0.175]{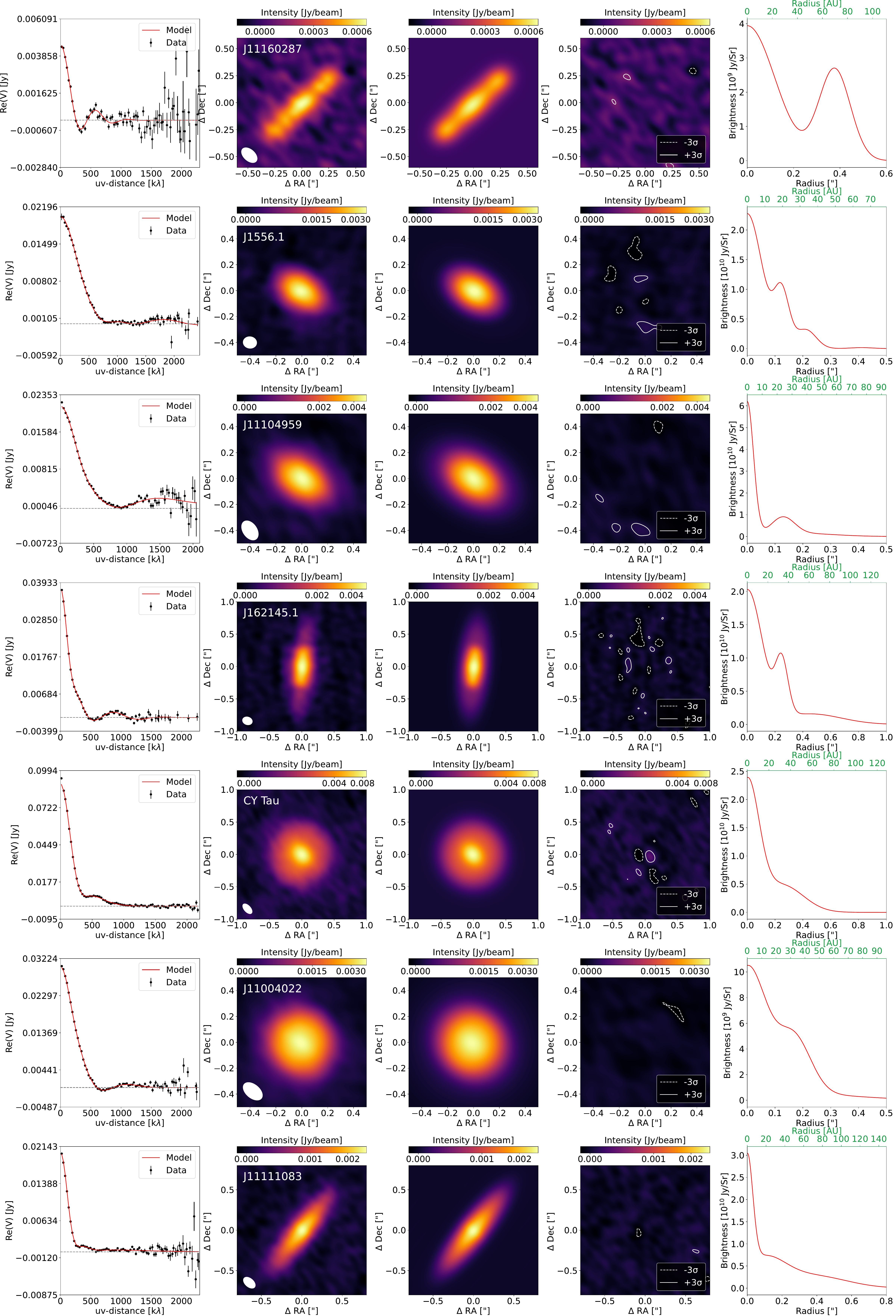}
    \caption{Same as for Figure \ref{fig:galario_first}, but for next seven modelled single discs.}
    \label{fig:galario_second}
\end{figure*}

\begin{figure*}[ht]
    %\hspace{-12mm}
    \centering
    \includegraphics[scale=0.175]{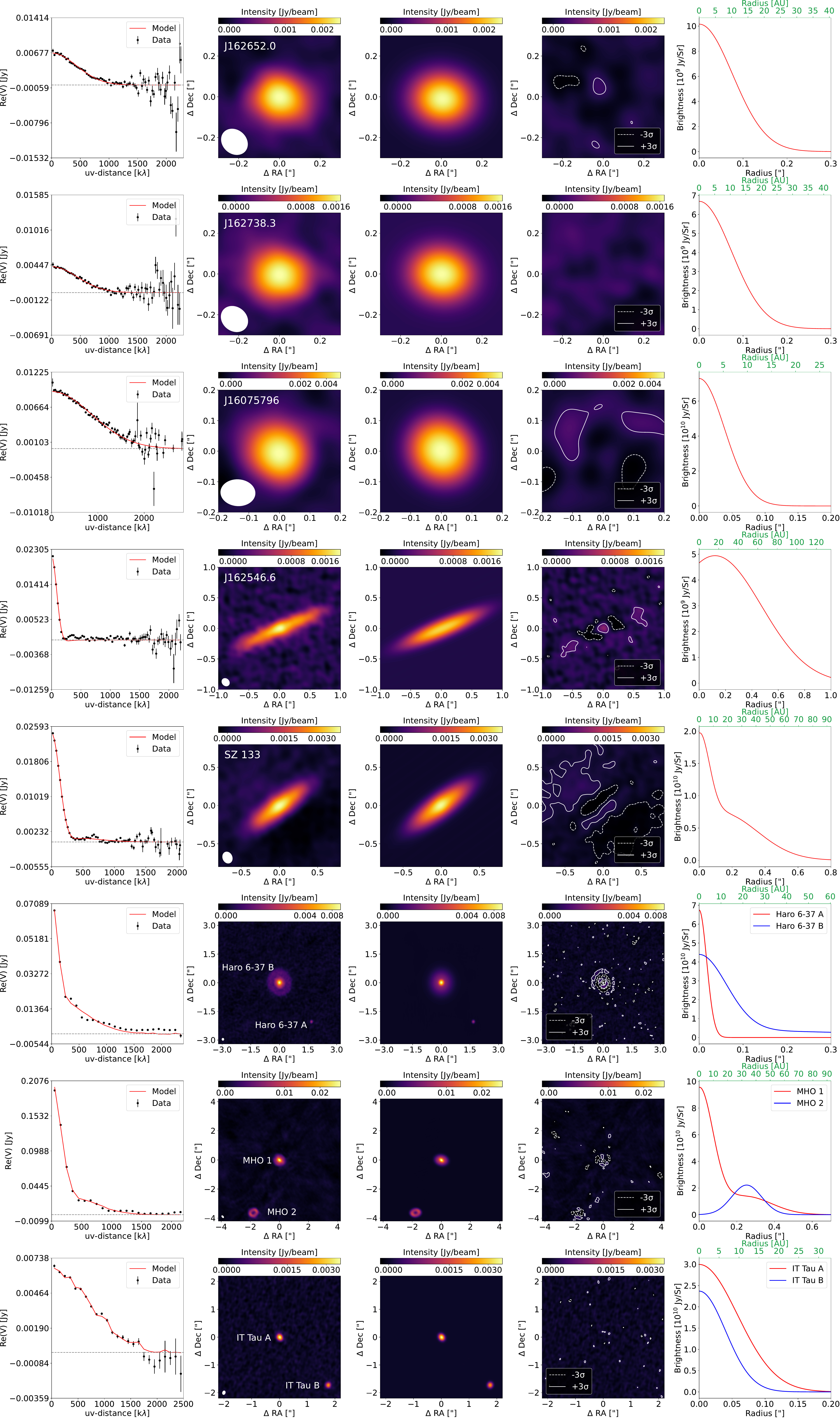}
    \caption{Same as for Figure \ref{fig:galario_first}, but for last five modelled single discs and the three modelled binary discs.}
    \label{fig:galario_third}
\end{figure*}

This appendix presents the results of the \texttt{Galario} visibility modelling for the 25 discs. For each single disc, we show the deprojected visibilities, cleaned image, best-fit model image, residuals, and radial brightness profile. The models reproduce the observed axisymmetric features with residuals below $\sim$3$\sigma$, using the minimal number of Gaussian components required. For the three binary systems, the results for both components are shown within the same figure: the visibilities represent the combined system, while the radial brightness profiles display separate curves for each disc. Discs are grouped by morphological classification: transition, ring, shoulder, compact, and highly-inclined. Figures \ref{fig:galario_first}, \ref{fig:galario_second}, and \ref{fig:galario_third} provide the model comparisons for each class.

\section{Upper Sco} %Nienke
\label{appendix:uppersco}
\begin{table*}[ht]
    \centering
    \caption{New Upper Sco sample.}
    \begin{tabular}{lccccccccl}
    \hline
      \textbf{Name}&\textbf{\textit{d}}&\textbf{\textit{F$_{\textbf{mm}}$}}&\textbf{Band}&\textbf{\textit{M$_{\textbf{dust}}$}}&\textbf{SpT}&\textbf{\textit{M$_{*}$}}&\textbf{\textit{R$_{\textbf{68\%}}$}}&\textbf{Disc type}&\textbf{Ref.$^\textbf{a}$} \\
      &\textbf{[pc]}&\textbf{[mJy]}&&\textbf{[M$_{\oplus}$]}&&\textbf{[M$_{\odot}$]}&\textbf{[AU]}&&\\
    \hline
    \end{tabular}
    \label{tab:uppsco}
    \caption*{$^a$ The first reference is for the stellar information, the second (and third) for the ALMA information.\\
    Full table available online.}
\end{table*}
The Upper Sco targets in the sample of \citet{NienkeGijs} was largely based on the first ALMA survey of this region by \cite{Barenfeld} with the addition of a handful of early type stars, resulting in a total of 72 Class II discs. However, \citet{Luhman2018} identified hundreds of new members in Upper Sco, later refined in \citet{Luhman2020} with \emph{Gaia DR2} data. \citet{Carpenter2025} presented a new ALMA Band 7 survey of Upper Sco based on the newly identified members at moderate resolution ranging from 0.1 to 0.3", and constrained disc sizes with visibility modelling for the detected discs. Based on these studies, we redefine the Upper Sco sample from \citet{NienkeGijs} with all protoplanetary discs, i.e. all members classified as either 'full' or 'transitional' discs by \citet{Luhman2020}, resulting in a total of 205 targets. The majority of these targets were included in \citet{Carpenter2025}, so fluxes and estimates on disc sizes are readily available. For a subset of 10 of the brightest targets, high-resolution ALMA data at 0.03" resolution were taken in program 2022.1.00646.S (PI Feng Long). For the 24 targets that were not included in \citet{Carpenter2025}, published ALMA Band 6 or 7 data were available for 10 targets, as well as ALMA archival data for seven targets. For seven targets, no ALMA data are available, which is less than 4\% of the total sample. We note that some of the newly identified Upper Sco members were previously considered as members of Ophiuchus, and included in ALMA Ophiuchus surveys. 

Using the ALMA data described above, we classify all discs based on visual inspection as transition disc (15), ring disc (4), highly-inclined disc (2), non-structured (129), non-detection (45) and unknown (10), similar to the approach in \citet{NienkeGijs}. The unknown category includes the seven discs without ALMA data as well as three discs with only very low resolution ALMA data of >0.5" (HD 144432, HD 145718 and PDS 415). We note that these three are all Group II Herbig discs which are generally very compact in millimetre emission \citep{Stapper}. 

For the non-structured discs, we use the estimated 68\% dust disc radii to classify the discs as either compact (<30 AU) or extended (>30 AU): only six discs are considered extended with radii between 31 and 39 AU, without further information on their substructure. Further inspection of these six targets shows that they all have small mm-fluxes <5 mJy and that the uncertainties on their derived sizes are relatively large: therefore, we classify them as compact as well.

The full disc sample of Upper Sco is presented in Table \ref{tab:uppsco}, including their mm-flux, disc radius,  classification and stellar information. The latter is taken from \citet{Barenfeld,Manara2020,Luhman2020} and references within. As \citet{Luhman2020} did not derive stellar masses, they were estimated from the spectral type based on the stellar masses of other Upper Sco members with the same spectral type.

\end{appendix}
\end{document}